\documentclass[pre,onecolumn,letterpaper,superscriptaddress,showpacs,floatfix]{revtex4}
\usepackage{amssymb,amsmath,bm}
\usepackage[dvips]{graphicx}
\usepackage[dvips]{color}

\begin{document}

\title{Global phase diagrams of binary dipolar fluid mixtures}

\date{\today}

\author{I. Szalai}
\email{szalai@almos.vein.hu}
\affiliation{Department of Physics, University of Veszpr\'em, 
H-8201 Veszpr\'em, POBox 158, Hungary} 
\author{S. Dietrich}
\email{dietrich@mf.mpg.de}
\affiliation{Max-Planck-Institut f\"ur Metallforschung, Heisenbergstr. 3,
D-70569 Stuttgart, Germany, and}
\affiliation{
Institut f\"ur Theoretische und Angewandte Physik,
Universit\"at Stuttgart, Pfaffenwaldring 57, D-70569 Stuttgart, Germany}

\begin{abstract}

We apply a modified mean-field density functional theory to determine the phase 
behavior of binary mixtures of Stockmayer fluids whose spherical constituents interact 
according to Lennard-Jones (LJ) pair potentials with embedded pointlike dipole moments. 
On the basis of systematic numerical calculations we construct the global phase diagrams 
of these systems in the three-dimensional  thermodynamic space of temperature, pressure, 
and chemical potential difference of the two components. The vapor-liquid, isotropic 
liquid -- isotropic liquid, isotropic liquid -- ferromagnetic liquid, 
and ferromagnetic liquid -- ferromagnetic liquid first-order phase 
separations are investigated. The loci of the second-order
isotropic fluid -- ferromagnetic fluid transition are calculated from Landau theory. 
Liquid-vapor and liquid-liquid critical lines, 
tricritical lines, triple lines, and lines of critical end points of the
binary Stockmayer mixtures are also determined. We discuss how the topology of the phase 
diagrams change upon varying the strengths of the two dipole moments of the two species 
as well as their sizes.

\end{abstract}

\pacs{64.10.+h, 64.75.+g, 68.18.Jk, 64.70.-p}

\maketitle

\section{Introduction}

There are various classes of fluids whose structural and thermodynamical properties are 
strongly influenced by the presence of either magnetic or electric dipolar interactions 
between their constituents: polar molecular liquids, dipolar liquid crystals, ferrofluids,
and magneto- or electrorheological fluids. These systems respond to external electric or 
magnetic fields and some may even exhibit long-ranged spontaneous orientational order 
in the liquid phases. The strongly anisotropic and long-ranged character of the dipolar 
forces causes particular difficulties for the theoretical description of these systems. 
By using simulations, integral equation theories, density functional theories, and phenomenological 
approaches significant progress has been achieved in understanding one-component dipolar 
fluids (see, e.g., Refs. \cite{teix1} and \cite{groh1} and references therein) albeit several 
aspects remain still under debate \cite{tlusty}.

If isotropic phases of dipolar fluids form interfaces (such as liquid-vapor or fluid-wall 
interfaces) the broken translational invariance gives rise to interesting local orientational
order tied to the interface \cite{egg,teix2,frodl,dang,enders}. Recently such
orientational profiles have become experimentally accessible through
ellipsometry \cite{muk1,muk2,cho1,cho2}. In these experiments the
orientational order at (critical) liquid-liquid \cite{muk1,muk2} and at
(noncritical) fluid-vapor \cite{cho1,cho2} interfaces of several binary
mixtures of polar (A) and nonpolar (B) components have been studied. In
particular the growing spatial extension of orientational order normal to the
interface and proportional to the bulk correlation length upon approaching the
continuous demixing transition of A-rich and B-rich phases in the bulk has
been analysed. These data have been discussed in terms of universal scaling
functions. Various forms of their shape have been proposed \cite{muk1,muk2},
including the occurrence of a surface correlation length different from the
bulk correlation length \cite{cho1,cho2}. These surprising interpretations
lack systematic theoretical analyses, which are particularly important because
the ellipsometry data provide only indirect access to orientational profiles
so that theoretical guidance is required. It is our goal to provide such
systematic theoretical studies for the aforementioned interfacial phenomena
near continuous demixing and to extend them to wetting phenomena. As a
necessary first step the present work aims at providing an overview of the
global phase behavior of polar binary liquid mixtures. This is an
indispensable prerequisite for studying the interfacial properties in a second
step. Moreover, in view of the numerous interesting features even of bulk
properties of dipolar fluids \cite{teix1,groh1,tlusty}, the analysis of
dipolar mixtures is also important in its own right.

Already some time ago there was an interesting first step towards the  
understanding of interfacial properties of dipolar liquid mixtures. 
In a combined experimental and theoretical study the liquid-vapor surface
tensions of two polar binary liquid mixtures have been studied for various 
temperatures and as function of  composition \cite{teix3}. Using density functional
theory (DFT) also relative adsorption as well as concentration and
orientational profiles have been calculated \cite{teix3}.  By resorting to an 
improved DFT it is our intention to expand the above analysis towards focusing 
on liquid-liquid demixing transitions, the formation of phases with
long-ranged orientational order, critical phenomena, and wetting phenomena 
and thus reaching out to the aforementioned more recent experimental studies.

There is a sizeable body of theoretical research available which  is devoted 
to the study of bulk properties of polar liquid mixtures. This encompasses
Monte Carlo simulations \cite{brad,chan,leeu1,jiang,gao,blair,cabral,kristof,kristof2,sadus},
molecular dynamics simulations 
\cite{leeu1,leeu2,moo,muller1,muller2,kriebel1,wang},
integral equations \cite{morris,lee,chen1,chen2}, 
cluster aggregation models \cite{troh,zubar,kant1,kant2,ivanov2}, construction 
of equations of state \cite{kriebel2}, thermodynamic perturbation theory 
\cite{woj,boda,valisko,ivanov},
structure formation in external fields \cite{islam}, and van der Waals and Ising 
fluids \cite{fenz}. There are few theoretical studies concerning the structure 
formation of dipolar mixtures at interfaces such as DFT and Monte Carlo 
studies at charged walls \cite{sena1,sena2} and integral equation approaches for 
dipolar mixtures in random confined geometries \cite{spol,ferna}. Beyond the
aforementioned study in Ref. \cite{teix3} the structure of dipolar mixtures at 
interfaces between fluid phases has been investigated by Monte Carlo
simulations of the liquid-liquid benzene-water interface \cite{linse} and by molecular 
dynamics simulations \cite{matsu1} and lattice-gas models \cite{matsu2} of the liquid-vapor
interface of water-methanol mixtures. In the context of this body of
research the present study is focused on obtaining global phase diagrams
using an appropriate version of DFT which opens the possibility for
interfacial studies later on.

Models for dipolar fluids have to capture three basic features of the pair
potentials between the fluid particles: short-ranged repulsion (which in
general is anisotropic reflecting molecular shapes), long-ranged attractive
dispersion forces (containing an omnipresent isotropic contribution $\sim {r^{-6}}$ at large
distances $r$ between the particles plus higher order anisotropic terms), and the
dipolar interaction $\sim {r^{-3}}$.

As long as one is not aiming for a quantitative description of a specific
system but for general phenomena and trends the so-called Stockmayer model has
turned out to be rather useful (see Refs. \cite{teix1} and \cite{groh1} and references
therein). It considers spherical particles interacting with Lennard-Jones (LJ) 
potentials plus pointlike permanent dipoles at the center. Such binary systems 
are characterized by three thermodynamic parameters (temperature $T$   
and two chemical potentials $\mu_A$  and $\mu_B$  for the two species $A$  and
$B$), three LJ interaction 
energy parameters $\epsilon_{ij}$ (${i,j=A,B}$), three LJ molecular diameter
parameters $\sigma_{ij}$, and two dipole moments $m_A$  and $m_B$ . 
After choosing an energy scale for $k_BT={\beta}^{-1}$  and the $\mu_i$  among 
the $\epsilon_{ij}$  and 
a length scale for the number densities among the $\sigma_{ij}$  one is left with a
six-dimensional interaction parameter space. By using DFT we analyze 
the three-dimensional ($T,p,\Delta\mu$) phase diagrams (where $p$ is the
pressure of the system and $\Delta\mu=\mu_B-\mu_A$) as they change
along certain paths within suitable subspaces of this interaction parameter space.

This Stockmayer model provides an approximate model for dipolar molecular
fluids as well as an effective model for ferrofluids with the solvent degrees 
of freedom taken into account via {\em effective} interaction energies
$\epsilon_{ij}$. Ferrofluids are often polydisperse \cite{kristof,huke}. 
In this respect the study of binary Stockmayer models 
provides also insight into the effects of polydispersity for ferrofluids in 
the special case of bimodal distributions of their sizes or of their magnetic 
moments.
 
Whereas one-component fluids with simple, radially symmetric pair
potentials give rise to only two fluid phases, i.e., liquid and vapor, binary 
fluids exhibit typically three fluid phases: vapor, $A$-rich liquid, and $B$-rich
liquid. The latter two emerge via phase segregation from a homogeneous liquid 
upon lowering the temperature. There is a line of critical points (with high
temperatures) limiting a two-dimensional sheet of first-order liquid-vapor
phase transitions, a line of critical points (at intermediate temperatures) limiting 
a two-dimensional sheet of first-order demixing transitions into $A$-rich and
$B$-rich liquid, and a triple line of phase coexistence between all three fluid
phases formed as the intersection of the two aforementioned sheets; at high 
temperatures the triple line ends at a critical end point where the line of 
critical demixing points meets the sheet of liquid-vapor coexistence 
(see, e.g., Ref. \cite{dietrich1} and references therein). At low temperatures freezing 
sets in such that there the triple line ends in a tetra point where the vapor 
phase, $A$-rich liquid, $B$-rich liquid, and the solid phase coexist. (For the
topology of this phase diagram see Ref. \cite{dietrich2}.) In the present study we focus 
on sufficiently high temperatures so that freezing is not of concern. (For
freezing in one-component dipolar fluids see Ref. \cite{groh1} and references
therein. For quasicrystalline order and glasslike structure formation in two-dimensional 
binary dipolar systems see Refs. \cite{scheffler,varga,mangold}.) 
The introduction of dipole moments shifts these phase boundaries 
and may even give rise to the formation of new liquid phases with 
spontaneous long-ranged orientational order.

In a recent series of papers \cite{range1,range2,range3} Range and Klapp have also extended 
the modified mean-field density functional theory to study the phase behavior of binary 
dipolar mixtures. There they have focused on the investigation 
of the isotropic fluid -- ferromagnetic fluid phase separations. Studying dipolar 
hard sphere mixtures (and briefly mentioning binary Stockmayer fluid mixtures 
in Ref. \cite{range2}) they have found that the isotropic fluid -- ferromagnetic 
fluid phase transitions are shifted towards higher densities as compared with the 
corresponding one-component dipolar fluids. These general trends are in agreement with 
our findings for the corresponding phase transitions in binary Stockmayer fluid mixtures. 
However, there are important differences to the case of dipolar hard sphere systems, 
because due to the additional presence of isotropic dispersion forces Stockmayer fluids 
exhibit also a vapor phase, which is indispensable for studying the interfacial 
properties mentioned at the beginning of the Introduction. These dispersion forces 
modify also the other phase transitions. Therefore in the following we explore the 
wider phase behavior of binary Stockmayer fluid mixtures. In Sec. \ref{dft}  we 
formulate the DFT for Stockmayer fluid mixtures. Our results are 
presented in Sec. III and summarized and discussed in Sec. IV.


\section{Density functional theory for Stockmayer fluid mixtures}
\label{dft}
\subsection{Microscopic model}

We study binary Stockmayer fluids which consist of spherically symmetric
particles interacting via Lennard-Jones (LJ) potentials with parameters 
$\sigma_{ij}$ and $\epsilon_{ij}$ (${i,j=A,B}$):
\begin{equation}
w_{ij}^{LJ}(r_{12})=4\epsilon_{ij}\left[\left(\frac{\sigma_{ij}}{r_{12}}\right)^{12}
-\left(\frac{\sigma_{ij}}{r_{12}}\right)^6\right].
\label{ljpot}
\end{equation}
In the following for the unlike LJ size parameter we assume 
$\sigma_{ij}=(\sigma_{ii}+\sigma_{jj})/2$. In addition there is the dipolar
interaction due to point dipoles embedded at the centers of spheres of
diameter $\sigma_{ii}$:
\begin{equation}
w_{ij}^{dip}(\bm{r}_{12},\omega_1,\omega_2)= -\frac{m_im_j}{r_{12}^3}
\left[\frac{[\widehat{\bm{m}}_1(\omega_1)\cdot\bm{r}_{12}]
[\widehat{\bm{m}}_2(\omega_2)\cdot\bm{r}_{12}]}{r_{12}^2}
-\left[\widehat{\bm{m}}_1(\omega_1)\cdot\widehat{\bm{m}}_2(\omega_2)\right]\right]
\Theta(r_{12}-\sigma_{ij}) 
\label{dippot}
\end{equation}
where particle 1 (2) of type $i$ ($j$) is located at $\bm{r}_1$ ($\bm{r}_2$) and
carries a dipole moment of strength $m_i$ ($m_j$) with an orientation given by
the unit vector $\widehat{\bm{m}}_1(\omega_1)$
($\widehat{\bm{m}}_2(\omega_2)$) with polar angles $\omega_1=(\theta_1,\phi_1)$
($\omega_2=(\theta_2,\phi_2)$); $\bm{r}_{12}=\bm{r}_1-\bm{r}_2$ is the
difference vector between the centers of the particle $i$ and the particle
$j$ with $r_{12}=|\bm{r}_{12}|$ and $\Theta(r)$ is the Heaviside step function. 
Thus the total pair potential is given by
\begin{equation}
w_{ij}(\bm{r}_{12},\omega_1,\omega_2)=w_{ij}^{LJ}(r_{12})
+w_{ij}^{dip}(\bm{r}_{12},\omega_1,\omega_2).
\label{totpot}
\end{equation}
In order to apply density functional theory, following Barker and Henderson
\cite{barker} this interaction potential is decomposed into a short-ranged
repulsive reference part
\begin{equation}
w_{ij}^{ref}(r_{12})=\Theta(\sigma_{ij}-r_{12})w_{ij}(\bm{r}_{12},\omega_1,\omega_2)=
\Theta(\sigma_{ij}-r_{12})w_{ij}^{LJ}(r_{12})
\label{refpot}
\end{equation}
and into a long-ranged attractive excess part
\begin{equation}
w_{ij}^{exc}(\bm{r}_{12},\omega_1,\omega_2)=
\Theta(r_{12}-\sigma_{ij})w_{ij}(\bm{r}_{12},\omega_1,\omega_2).
\label{atrpot}
\end{equation}
This decomposition allows one to choose a suitable reference system (a hard
sphere system with temperature dependent diameters) for which reliable
approximations for the free energy are known. The excess part
$w_{ij}^{exc}(\bm{r}_{12},\omega_1,\omega_2)$ is treated perturbatively in a
suitable way (see below).

\subsection{Modified mean-field density functional theory}

Our analysis is based on the following grand canonical variational functional,
which is the binary extension of the one component functional used in
Refs. \cite{groh2} and \cite{groh3}:
\begin{equation}
\frac{\Omega}{V}=f_{HS}^{ref}(\rho_A,\rho_B,T)+\beta^{-1}\sum_{i=A,B}\rho_i\int
d\omega\alpha_i(\omega)\ln[4\pi\alpha_i(\omega)]+\frac{\Omega_{int}}{V}+
\sum_{i=A,B}\rho_i\mu_i ,
\label{func1}
\end{equation}
where $V$, $T$, and $\mu_i$ denote the volume of the fluid, its temperature,
and the chemical potential of the $i$th component, respectively,
$\beta=1/(k_BT)$ with the Boltzmann constant $k_B$, $\rho_i=N_i/V$ is the number
density of the species $i$ taken to be spatially constant for the present bulk
study. $\alpha_i(\omega)$ is the orientational probability density
distribution (i.e., $\rho_i(\omega)=\rho_i\alpha_i(\omega)$ with
$\int d\omega\alpha_i(\omega)=1$ is the number density of particles of type
$i$ with orientation $\omega$). The first term in Eq. (\ref{func1}) is the
free energy density of the reference system due to Mansoori et
al. \cite{mansoori}:
\begin{equation}
f_{HS}^{ref}(\rho_A,\rho_B,T)=\beta^{-1}\sum_{i=A,B}\rho_i[\ln(\Lambda_i^3\rho_i)-1]
+\beta^{-1}\frac{6}{\pi}\left[\left(\frac{\xi_2^3}{\xi_3}-\xi_0\right)\ln(1-\xi_3)
+\frac{3\xi_1\xi_2}{1-\xi_3}+
\frac{\xi_2^3}{\xi_3(1-\xi_3)^2}\right]
\label{manseq}
\end{equation}
where $\Lambda_i$ denotes the thermal de Broglie wavelength of the particles
$i$ and
\begin{equation}
\xi_k=\frac{\pi}{6}\sum_{i=A,B}\rho_i(d_{ii})^k,\hspace{.5in}k=0,1,2,3\hspace{.2in},
\label{xik}
\end{equation}
with temperature dependent hard sphere diameters as given by Barker and
Henderson \cite{barker}: 
\begin{equation}
d_{ii}(T)=\int_{0}^{\sigma_{ii}}dr_{12}(1-\exp(-\beta w_{ii}^{ref}(r_{12}))). 
\label{bh}
\end{equation}
For $\sigma_{AA}=\sigma_{BB}$ Eq. (\ref{manseq}) reduces to the
Carnahan-Starling formula. In  Eq. (\ref{func1}) the second contribution to
the grand canonical functional takes into account the loss of entropy if the
orientational distributions of the particles are not isotropic. For the
interaction contribution due to the long-ranged part of the pair potential
between the spherical particles with embedded point dipoles we adopt
\begin{equation}
\frac{\Omega_{int}}{V}=\frac{1}{2\beta}\sum_{i,j=A,B}\rho_i\rho_j
\int d^3r_1\int d^3r_2\int d\omega_1\int d \omega_2 \alpha_i(\omega_1)
\alpha_j(\omega_2)\exp(-\beta w_{ij}^{ref}(r_{12}))
f_{ij}^{exc}(\bm{r}_{12},\omega_1,\omega_2)
\label{funcint}
\end{equation}
where
\begin{equation}
f_{ij}^{exc}(\bm{r}_{12},\omega_1,\omega_2)=1-
\exp(-\beta w_{ij}^{exc}(\bm{r}_{12},\omega_1,\omega_2)) 
\label{mayer}
\end{equation}
are the Mayer functions with the excess interaction potential. Equation
(\ref{funcint}) amounts to using the low-density approximation for the pair 
distribution functions:
\begin{equation}
g_{ij}(r_{12})=\exp(-\beta w_{ij}^{ref}(r_{12})). 
\label{paircorr}
\end{equation} 
For uniaxial molecules the orientational distribution functions
$\alpha_i(\omega)$ depend only on the angle $\theta$ and thus allow for the
following expansion into Legendre polynomials:
\begin{equation}
2\pi\alpha_i(\omega)=\bar{\alpha}_i(\cos\theta)=\sum_{l=0}^{\infty}
\alpha_i^{(l)}P_l(\cos\theta) 
\label{orient}
\end{equation}
where  
\begin{equation}
\alpha_i^{(l)}=\frac{2l+1}{2}\int_{-1}^{1}dx\bar{\alpha}_i(x)P_l(x).
\label{alfai}
\end{equation}
Following Refs. \cite{groh2,groh3} in order to calculate the spatial integrals
in Eq. (\ref{funcint}) we consider a finite volume $V$ which has the shape of
a prolate rotational ellipsoid with aspect ratio $k$ ($k\geq1$). The Mayer
functions can be expressed in terms of the rotational invariants
$\Phi_{l_1l_2l}$ (for the definition of these functions see, e.g.,
Ref. \cite{groh3}) so that the corresponding integrals yield
\begin{equation}
\frac{\Omega_{int}}{V}=\sum_{i,j=A,B}\rho_i\rho_j
\sum_{l=0}^{\infty}u_{ij}^{(l)}\alpha_i^{(l)}\alpha_j^{(l)}
\label{intalf}
\end{equation}
where
\begin{equation}
u_{ij}^{(l)}=-\frac{8\pi}{9}\delta_{l,1}I(k)m_im_j
-\beta^{-1}\frac{4\sqrt{\pi}(-1)^l}{(2l+1)^{3/2}}
\int_{\sigma_{ij}}^{\infty}dr_{12}r_{12}^2\int d\omega_1d\omega_2f_{ij}^{exc}
(r_{12},\omega_1,\omega_2,\omega_{12})
\Phi_{ll0}^{\star}(\omega_1,\omega_2,\omega_{12}).
\label{uij}
\end{equation}
Further details of this expansion can be found in Ref. \cite{groh3}. The
function $I(k)$ describes the shape dependence of the grand canonical
potential, which enters only via the coefficients $u_{ij}^{(1)}$ and is given
as
\begin{equation}
I(k)=\frac{k^2+2}{3(k^2-1)}-\frac{k}{(k^2-1)^{3/2}}
\ln \left(k+\sqrt{k^2-1}\right).
\label{ik}                                  
\end{equation}
Therefore the grand canonical functional reads
\begin{equation}
\frac{\Omega}{V}=f_{ref}^{HS}(\rho_A,\rho_B,T)+k_BT\sum_{i=A,B}\rho_i\int_{-1}^1
dx\bar{\alpha}_i(x)\ln[2\bar{\alpha}_i(x)]+\sum_{i,j=A,B}\rho_i\rho_j
\sum_{l=0}^{\infty}u_{ij}^{(l)}\alpha_i^{(l)}\alpha_j^{(l)}-\sum_{i=A,B}\rho_i\mu_i .
\label{grandf}                                  
\end{equation}
The equilibrium configuration for given $T$, $\mu_A$, and $\mu_B$ is
determined by minimizing the total grand canonical functional 
with respect to $\rho_A$,
$\rho_B$, and the functions $\alpha_A(x)$ and $\alpha_B(x)$:
\begin{equation}
\frac{1}{V}\left.\frac{\partial \Omega}{\partial \rho_i}\right|_{\rho_{j\neq
    i},\alpha_i(x)}=0,\hspace{.5in}i=A,B
\label{min1}                                  
\end{equation}
\begin{equation}
\frac{1}{V}\left.\frac{\delta \Omega}{\delta \alpha_i(x)}\right|_{\alpha_{j\neq
    i}(x),\rho_i}=0,\hspace{.5in}i=A,B .
\label{min2}                                  
\end{equation}
Equations (\ref{grandf}) and (\ref{min1}) lead to
\begin{equation}
\mu_i^{HS}(\rho_A,\rho_B,T)+k_BT\int_{-1}^1
dx\bar{\alpha}_i(x)\ln[2\bar{\alpha}_i(x)]
+2\sum_{j=A,B}\rho_j\sum_{l=0}^{\infty}u_{ij}^{(l)}\alpha_i^{(l)}\alpha_j^{(l)}-\mu_i=0,
\hspace{.5in}i=A,B.
\label{mu1}
\end{equation}
Equation (\ref{min2}) yields
\begin{equation}
\bar{\alpha}_i(x)=C_i\exp\left(-\beta\sum_{l=1}^{\infty}(2l+1)P_l(x)
\sum_{j=A,B}\rho_ju_{ij}^{(l)}\alpha_j^{(l)}\right),
\hspace{.5in}i=A,B ,
\label{alix}                                  
\end{equation}
with
\begin{equation}
\frac{1}{C_i}=\int_{-1}^1dx \exp\left(-\beta\sum_{l=1}^{\infty}(2l+1)P_l(x)
\sum_{j=A,B}\rho_ju_{ij}^{(l)}\alpha_j^{(l)}\right),
\hspace{.5in}i=A,B .
\label{cix}                                  
\end{equation}
For the second term in Eq. (\ref{grandf}) one finds after some elementary
calculations:
\begin{equation}
\beta^{-1}\int_{-1}^1 dx\bar{\alpha}_i(x)\ln[2\bar{\alpha}_i(x)]=
k_BT\ln(2C_i)-2\sum_{j=A,B}\rho_j\sum_{l=1}^{\infty}
u_{ij}^{(l)}\alpha_i^{(l)}\alpha_j^{(l)},
\hspace{.5in}i=A,B.
\label{ori1}                                  
\end{equation}
Equations (\ref{alfai}) and (\ref{alix}) lead to a system of coupled nonlinear
equations for the coefficients $\alpha_A^{(l)}$ and $\alpha_B^{(l)}$:
\begin{equation}
\alpha_i^{(l)}=\frac{2l+1}{2}\frac{\int_{-1}^1dx
P_l(x)\exp\left(-\beta\sum_{l=1}^{\infty}(2l+1)
P_l(x)\sum_{j=A,B}\rho_ju_{ij}^{(l)}\alpha_j^{(l)}\right)}{\int_{-1}^1dx
\exp\left(-\beta\sum_{l=1}^{\infty}(2l+1)P_l(x)\sum_{j=A,B}\rho_ju_{ij}^{(l)}\alpha_j^{(l)}
\right)},
\hspace{.5in}i=A,B.
\label{ali}                                  
\end{equation}
Equation (\ref{mu1}) together with Eqs. (\ref{cix}) and (\ref{ori1}) give the following 
expressions for the chemical potentials of the two components: 
\begin{equation}
\mu_i=\mu_i^{HS}+k_BT\ln(2C_i)+\frac{1}{2}\sum_{j=A,B}\rho_ju_{ij}^{(0)},
\hspace{.5in}i=A,B.
\label{kem}                                  
\end{equation}
The chemical potentials $\mu_A$ and $\mu_B$ can be eliminated from Eq. (\ref{grandf})
and therefore $\Omega$ can be expressed as
\begin{equation}
\frac{\Omega}{V}=f_{ref}^{HS}(\rho_A,\rho_B,T)-\sum_{i=A,B}\rho_i\mu_i^{HS}
(\rho_A,\rho_B,T)-\sum_{i,j=A,B}\rho_i\rho_j\sum_{l=0}^{\infty}
u_{ij}^{(l)}\alpha_i^{(l)}\alpha_j^{(l)}.
\label{ompot}                                  
\end{equation}
Due to $\Omega=-pV$ the corresponding pressure of an equilibrium phase as a
function of the densities and the temperature is 
\begin{equation}
p =-f_{ref}^{HS}(\rho_A,\rho_B,T)+\sum_{i=A,B}\rho_i\mu_i^{HS}
(\rho_A,\rho_B,T)+\sum_{i,j=A,B}\rho_i\rho_j\sum_{l=0}^{\infty}
u_{ij}^{(l)}\alpha_i^{(l)}\alpha_j^{(l)}.
\label{press}                                  
\end{equation}
These results show that the knowledge of the functions $u_{ij}^{(l)}(T)$ is 
necessary to obtain the coefficients $\alpha_i^{(l)}$ of the orientational 
distribution functions (Eq.(\ref{ali})) and to obtain the different thermodynamic 
functions (Eqs. (\ref{kem}-\ref{press})). In computing these functions we use the same 
approximations as in Ref. \cite{groh3}, i.e., we expand the Mayer functions in 
Eq. (\ref{uij}) for small dipole moments $m_i$ which allows us to perform the 
angular integrations analytically. As in Ref. \cite{groh3} we have determined 
the expansion coefficients up to $O(m^8)$. This gives a satisfactory accuracy 
for reduced dipole moments $m_i^{\star}=m_i/\sqrt{\epsilon_{AA}\sigma_{AA}^3} 
\lesssim2$ \cite{groh3}. The omission of the higher order terms implies a truncation in 
the summations at $l=4$ in all equations which contain such a summation over $l$. 
The corresponding coefficients $u_{ij}^{(l)}(T)$ are:
\begin{equation}
u_{ij}^{(0)}=\frac{8\pi}{\beta}\int_{\sigma_{ij}}^{\infty}dr_{12}r_{12}^2
\left(1-\exp(-\beta w_{ij}^{LJ}(r_{12})\right)
-\frac{8\pi\beta}{3\sigma_{ij}^3}m_i^2m_j^2I_4(\beta\epsilon_{ij})
-\frac{8\pi\beta^3}{25\sigma_{ij}^9}m_i^4m_j^4I_{10}(\beta\epsilon_{ij}),
\label{u0}                                  
\end{equation}
\begin{equation}
u_{ij}^{(1)}=-\frac{8\pi}{9}m_im_jI(k)-\frac{16\pi\beta^2}
{225\sigma_{ij}^6}m_i^3m_j^3I_7(\beta\epsilon_{ij}),
\label{u1}                                  
\end{equation}
\begin{equation}
u_{ij}^{(2)}=-\frac{8\pi\beta}{375\sigma_{ij}^3}m_i^2m_j^2I_4(\beta\epsilon_{ij})
-\frac{32\pi\beta^3}{6125\sigma_{ij}^9}m_i^4m_j^4I_{10}(\beta\epsilon_{ij}),
\label{u2}                                  
\end{equation}
\begin{equation}
u_{ij}^{(3)}=\frac{16\pi\beta^2}{25752\sigma_{ij}^6}m_i^3m_j^3I_7(\beta\epsilon_{ij}),
\label{u3}                                  
\end{equation}
and
\begin{equation}
u_{ij}^{(4)}=-\frac{8\pi\beta^3}{99225\sigma_{ij}^9}m_i^4m_j^4
I_{10}(\beta\epsilon_{ij})
\label{u4}                                  
\end{equation}
with
\begin{equation}
I_n(y)=\int_1^{\infty}dx x^{-n}\exp[4y(x^{-6}-x^{-12})].
\label{in}                                  
\end{equation}

\subsection{Phase equilibria and phase diagrams}

The phase diagrams at a given temperature $T$ follow from requiring the
equality of the chemical potentials of both components and of the pressure for
the densities $\rho_i^{(I)}$ and $\rho_i^{(II)}$ characterizing the coexisting
phases I and II:
\begin{eqnarray}
\left.\mu_A\right|_{\rho_i^{(I)}, \bar{\alpha}_i^{(I)}(x)}&=&
\left.\mu_A\right|_{\rho_i^{(II)}, \bar{\alpha}_i^{(II)}(x)},\nonumber\\
\left.\mu_B\right|_{\rho_i^{(I)}, \bar{\alpha}_i^{(I)}(x)}&=&
\left.\mu_B\right|_{\rho_i^{(II)}, \bar{\alpha}_i^{(II)}(x)},\label{caca}\\
\left.p\right|_{\rho_i^{(I)}, \bar{\alpha}_i^{(I)}(x)}&=&
\left.p\right|_{\rho_i^{(II)}, \bar{\alpha}_i^{(II)}(x)}.\nonumber                              
\end{eqnarray} 
The functions $\bar{\alpha}_i^{(I)}(x)$ and  $\bar{\alpha}_i^{(II)}(x)$ 
($i=A,B$) denote the corresponding equilibrium orientational distributions 
obtained from Eqs. (\ref{ali}) and (\ref{alix}). As mentioned in the 
Introduction we consider four kinds of two-phase equilibria: isotropic liquid 
-- isotropic vapor, isotropic A-rich liquid -- isotropic B-rich liquid, 
isotropic liquid -- ferromagnetic liquid, and ferromagnetic A-rich liquid -- 
ferromagnetic B-rich liquid. For binary fluids Gibbs' phase rule allows for 
three-phase coexistence lines and four-phase (tetra) coexistence points in the 
thermodynamic space ($T,p,\Delta\mu$). For isotropic phases the corresponding 
equations for the chemical potentials and the pressure are simpler because for 
them $\bar{\alpha}_i(x)=1/2$ (i.e., $\alpha_i^{(0)}=1/2$ and 
$\alpha_i^{(l)}=0$ if $l\geq1$).

In order to locate the surface of critical points separating the isotropic and
the ferromagnetic liquid phases by second-order phase transitions, similar as
in Ref. \cite{groh3} we expand the orientation dependent ideal gas
contributions to the interaction part $\Omega_{int}$ of the grand canonical
functional for small deviations from isotropy, i.e., for small
$\alpha_i^{(l)}$. Using the results for the one component dipolar 
fluid with $l=1$ the leading orientation dependent contribution 
to the grand canonical potential turns out as 
\begin{equation}
\frac{\Delta\Omega}{V}=\sum_{i,j=A,B}\left(\rho_i\rho_ju_{ij}^{(1)}
+\frac{2\rho_i}{3\beta}\delta_{ij}\right)\alpha_i^{(1)}\alpha_j^{(1)}+...
\label{sec}                                  
\end{equation}
where $\delta_{ij}$ is the Kronecker delta symbol. Obviously, for given
$\beta$, $\rho_A$, and $\rho_B$ the isotropic configuration minimizes the
grand canonical functional if the quadratic form defined by Eq. (\ref{sec}) is
positive definite. Provided $\rho_Au_{AA}^{(1)}+2/(3\beta)>0$, this quadratic
form is positive definite if and only if its determinant is positive
\cite{grad}. 
For all the cases studied it turns out that this inequality is satisfied. Thus
the zero of the determinant defines the surface which separates the isotropic
liquid phase from the ferromagnetic liquid phase:
\begin{equation}
\left(\rho_Au_{AA}^{(1)}+\frac{2}{3\beta}\right)
\left(\rho_Bu_{BB}^{(1)}+\frac{2}{3\beta}\right)-\rho_A\rho_B\left(u_{AB}^{(1)}\right)^2=0.
\label{sec2}                                  
\end{equation}
In the one-component limit Eq. (\ref{sec2}) reduces to the equation for the
critical line given in Ref. \cite{groh3}.  

\section{Results and discussion}

In the following we shall use reduced quantities: $T^{\star}=k_BT/\epsilon_{AA}$
as reduced temperature, $\rho^{\star}=(\rho_A+\rho_B)\sigma_{AA}^3$ as reduced
total density, $x_i=\rho_i/(\rho_A+\rho_B)$ as the concentration of species
$i$, $p^{\star}=p\sigma_{AA}^3/\epsilon_{AA}$ as reduced pressure,
$\mu_i^{\star}=(\mu_i-k_BT
\ln(\Lambda_i^3/\sigma_{AA}^3))/\epsilon_{AA}$ as the reduced chemical
potential of species $i$, and
$m_i^{\star}=m_i/\sqrt{\epsilon_{AA}\sigma_{AA}^3}$ as the reduced dipole
moments. For our numerical analysis we consider the following choices for the
Lennard-Jones potential parameters: $\epsilon_{BB}/\epsilon_{AA}=0.95$,
$\epsilon_{AB}/\epsilon_{AA}=0.75$, and $\sigma_{AA}=\sigma_{BB}=\sigma_{AB}$. 
The case $\sigma_{AA}\neq\sigma_{BB}$ will be discussed in Subsec. III C. 
The bulk phase diagrams of the binary dipolar fluid mixtures are calculated in 
the thermodynamic space spanned by temperature $T$, pressure $p$, and the 
difference of the chemical potentials $\Delta\mu=\mu_B-\mu_A$ of the two 
components. In order to obtain the phase boundaries in this space for fixed 
$T$ the corresponding phase equilibrium equations (Eq. (\ref{caca})) have to be 
solved simultaneously under the constraint 
\begin{equation} 
\Delta\mu-\mu_B(\rho_A^I,\rho_B^I,T)+\mu_A(\rho_A^I,\rho_B^I,T)=0. 
\label{kont}                                   
\end{equation} 
The pressure $p$ follows from Eq. (\ref{press}).

\subsection{Binary mixtures of particles with equal size and equal dipole moment}

\begin{figure}[t]
\vspace{0.5cm} 
\begin{center}
\rotatebox{0}{\scalebox{0.45}{\includegraphics{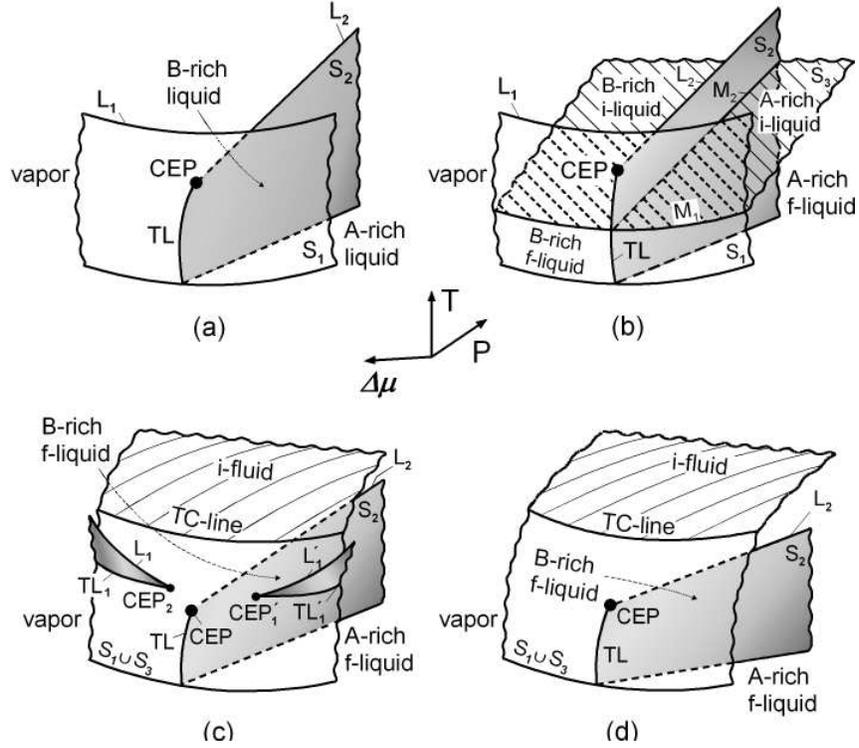}}}\\
\end{center}
\caption{Schematic bulk phase diagrams of binary Stockmayer fluid mixtures in
the thermodynamic space of temperature $T$, pressure $p$, and chemical
potential difference $\Delta\mu=\mu_B-\mu_A$ of the two components for various equal 
dipole moments. The phase diagram in (a) represents a non-polar 
($m_A^{\star}=m_B^{\star}=0$) binary fluid mixture which exhibits three 
phases: vapor, A-rich liquid, and B-rich liquid. They are separated by sheets $S_1$
and $S_2$ of first-order phase transitions, delimited by lines $L_1$ and
$L_2$, respectively, of second-order phase transitions. $S_1$ and $S_2$
intersect along the triple line $TL$ which ends at the critical end point
$CEP$. The phase diagram in (b) represents a dipolar
($m_A^{\star}=m_B^{\star}=1$) binary fluid mixture for which a third sheet
$S_3$ (hatched) appears, which separates the isotropic liquid from the
ferromagnetic liquid. $S_3$ is a sheet of second-order phase transitions. 
The intersection of the sheets $S_1$ and $S_3$ forms a line $M_1$ of critical
end points. The intersection of the sheets $S_2$ and $S_3$ forms another 
line $M_2$ of critical end points. Upon increasing the dipole moments 
($m_A^{\star}=m_B^{\star}=1.5$) in (c) the sheet $S_3$ has been raised above
$L_1$ and has merged with $S_1$ into the sheet $S_1\cup{S_3}$ with a tricritical
line ($TC$-line), which separates the first-order transition (lower part of
the sheet) from the second-order transition (hatched upper part). The
isotropic liquid-vapor transitions are reduced to two winglike surfaces,
which are connected to the sheet $S_1\cup{S_3}$ along the triple lines $TL_1$
and $TL_1^{\prime}$. These triple lines end in critical end points $CEP_1$ and
$CEP_1^{\prime}$. The liquid-vapor critical lines $L_1$ and $L_1^{\prime}$ are
the upper edges of the wings. Upon a further increase of the dipole moments
($m_A^{\star}=m_B^{\star}=2$) one obtains the phase diagram shown in (d) where
there are only two sheets: $S_2$ and $S_1\cup{S_3}$. The wings associated with
the isotropic liquid -- isotropic vapor transitions have disappeared and only
the isotropic vapor -- ferromagnetic liquid and the ferromagnetic A-rich liquid
-- ferromagnetic B-rich liquid phase transitions remain.}
\label{fig1}
\end{figure}

\begin{figure}[t]
\vspace{0.5cm} 
\begin{center}
\rotatebox{0}{\scalebox{0.4}{\includegraphics{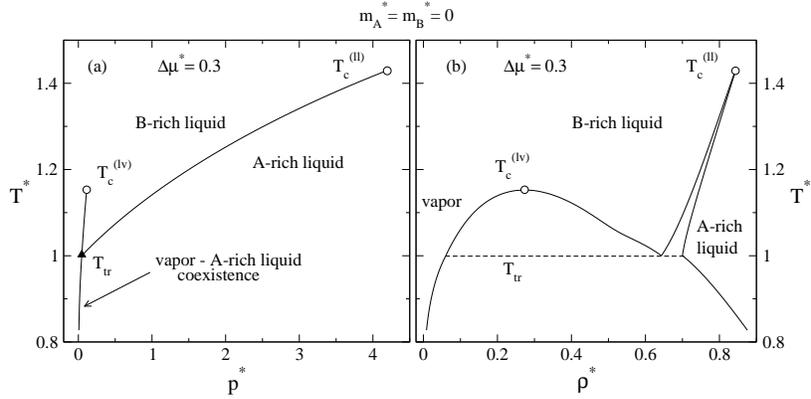}}}\\
\end{center}
\caption{Phase diagrams of a non-polar ($m_A^{\star}=m_B^{\star}=0$) binary 
Lennard-Jones fluid mixture in the pressure-temperature (a) and 
density-temperature (b) plane for $\Delta\mu^{\star}=0.3$; 
$\epsilon_{BB}/\epsilon_{AA}=0.95$, $\epsilon_{AB}/\epsilon_{AA}=0.75$, 
$\sigma_{AA}=\sigma_{BB}=\sigma_{AB}$. $T_{tr}$, $T_c^{(lv)}$, and 
$T_c^{(ll)}$ denote the triple, liquid-vapor critical, and liquid-liquid 
critical temperature, respectively. In (a) all lines are lines of first-order 
phase transitions. The vapor phase occurs only at very low pressure. In (b) 
the full curves enclose two-phase regions. The dashed line indicates three-phase 
coexistence. Below the triple point temperature the vapor coexists
with the A-rich liquid. Between $T_{tr}$ and $T_c^{(lv)}$ there are three
possible phases: the vapor, the isotropic B-rich liquid, and the A-rich
liquid. Between $T_c^{(lv)}$ and $T_c^{(ll)}$ there is only B-rich liquid -- 
A-rich liquid coexistence. At high densities the system freezes. The 
corresponding solid phases are not shown because they are not accessible by 
the present theory.}
\label{fig2} 
\end{figure}

As a first step we have determined the phase diagrams of non-polar binary 
Lennard-Jones fluid mixtures ($m_A^{\star}=m_B^{\star}=0$) for the aforementioned 
parameters. Based on our systematic calculations in the thermodynamic space 
($T,p,\Delta\mu$) these phase diagrams take on shapes as shown schematically in
Fig. \ref{fig1}(a). $S_1$ and $S_2$ are sheets of first-order phase
transitions separating the vapor phase from the fluid phase and the A-rich
liquid phase from the B-rich liquid phase, respectively. These sheets are
bounded by lines $L_1$ and $L_2$, respectively, of second-order phase
transitions marking the onset of liquid-vapor separation ($L_1$) and demixing 
($L_2$) upon lowering the temperature. The intersection of $S_1$ and $S_2$
forms the triple line $TL$ of three-phase coexistence ending at the critical
end point $CEP$ where $L_2$ hits $S_1$. This type of phase diagrams corresponds 
to type II in the classification scheme of phase diagrams of binary fluid 
mixtures given by Scott and van Konynenburg \cite{scott,konynen}. For the 
non-polar binary fluid mixture Fig. \ref{fig2} presents our numerical results 
for the pressure-temperature (Fig. \ref{fig2}(a)) and the density-temperature 
(Fig. \ref{fig2}(b)) phase diagrams for $\Delta\mu^{\star}=0.3$. At low 
temperatures ($T<T_{tr}$) the A-rich liquid coexists with the vapor.  
At the triple point temperature $T_{tr}^{\star}=1.003$ three phases coexist: 
vapor, B-rich liquid, and A-rich liquid. Above $T_{tr}$ up to the liquid-vapor
critical temperature ($T_c^{(lv)\star}=1.153$) there is coexistence between
the vapor and the B-rich liquid and between the A-rich liquid and the
B-rich liquid. In the temperature range $T_c^{(lv)}<T<T_c^{(ll)}$ with
$T_c^{(ll)\star}=1.429$ there is only coexistence between the A-rich liquid
and the B-rich liquid.

\begin{figure}[t]
\vspace{0.5cm} 
\begin{center}
\rotatebox{0}{\scalebox{0.4}{\includegraphics{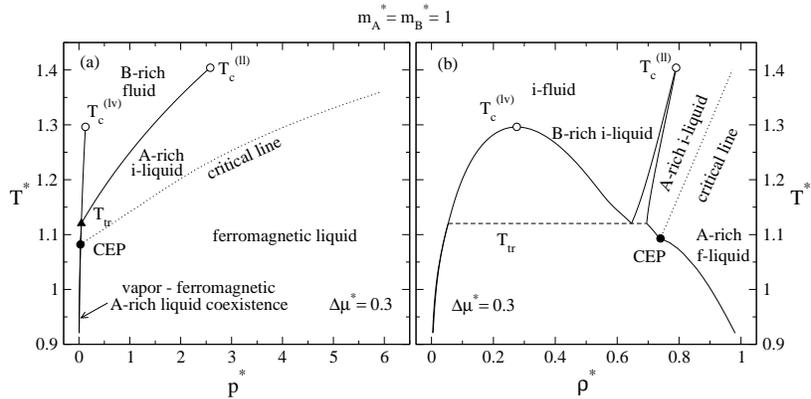}}}\\
\end{center}
\caption{Phase diagrams of a binary Stockmayer fluid mixture for
$m_A^{\star}=m_B^{\star}=1$ in the pressure-temperature (a) and
density-temperature (b) plane for $\Delta\mu^{\star}=0.3$. The LJ parameters
are the same as in Fig. \ref{fig2}. The dotted line denotes a line of
second-order phase transitions. It ends at a critical end point ($CEP$) where
the ferromagnetically critical liquid coexists with the non-critical
vapor. The critical line divides the A-rich liquid region into an isotropic
and ferromagnetic part. For more details see the caption of Fig. \ref{fig2}.}
\label{fig3}
\end{figure}

\begin{figure}[t] 
\vspace{0.5cm}
\begin{center}
\rotatebox{0}{\scalebox{0.4}{\includegraphics{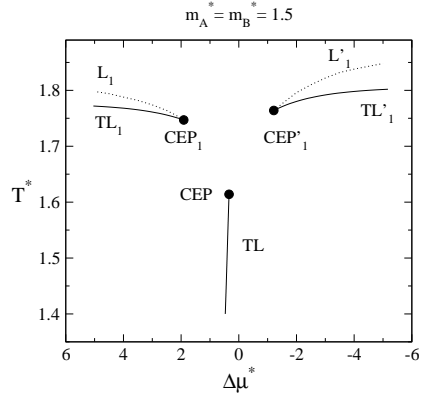}}}\\
\end{center}
\caption{Projections onto a plane $p=const$ of the triple lines and of the
liquid-vapor critical lines of a binary Stockmayer fluid mixtures with
$m_A^{\star}=m_B^{\star}=1.5$; the LJ parameters are the same as in
Figs. \ref{fig2} and \ref{fig3}. In accordance with Fig. \ref{fig1}(c) along 
$TL$ there is isotropic vapor -- A-rich
ferromagnetic liquid -- B-rich ferromagnetic liquid three phase
coexistence. This triple line ends at a critical end point ($CEP$). Both
$TL_1$ and $TL_1^{\prime}$ describe isotropic vapor -- isotropic fluid --
ferromagnetic fluid three-phase coexistence. These two triple lines end at
critical end points $CEP_1$ and $CEP_1^{\prime}$. The dashed lines $L_1$ and
$L_1^{\prime}$ represent the projections of the liquid-vapor critical lines
delimiting the upper end of the wings shown in Fig. \ref{fig1}(c). Note
that as in Fig. \ref{fig1} $\Delta\mu$ increases towards the left.}
\label{fig4}
\end{figure}

\begin{figure}[t] 
\vspace{0.5cm}
\begin{center}
\rotatebox{0}{\scalebox{0.4}{\includegraphics{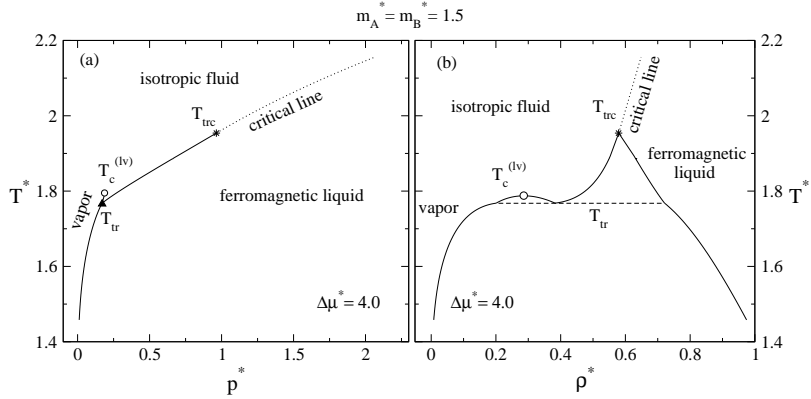}}}\\
\end{center}
\caption{Phase diagrams of a binary Stockmayer fluid mixture for 
$m_A^{\star}=m_B^{\star}=1.5$ (compare Fig. \ref{fig1}(c)) in the 
pressure-temperature (a) and density-temperature (b) plane for 
$\Delta\mu^{\star}=4$; the LJ parameters are the same as in 
Figs. \ref{fig2}-\ref{fig4}. Below the triple point temperature $T_{tr}$ the 
isotropic vapor coexists with the ferromagnetic liquid. Between $T_{tr}$ and 
$T_c^{(lv)}$ there are three possible phases: the isotropic vapor, the isotropic 
liquid, and the ferromagnetic liquid. The first-order phase transition between 
the isotropic fluid and ferromagnetic liquid turns into a second-order phase 
transition at the tricritical temperature $T_{trc}$. Above the tricritical 
temperature the dotted line represents the corresponding line of critical points. 
For the value of the 
chemical potential difference $\Delta\mu^{\star}=4$ there is no A-rich liquid 
-- B-rich liquid phase separation (i.e., the plane $\Delta\mu^{\star}=4$ does not 
intersect the sheet $S_2$; see Figs. \ref{fig1} and \ref{fig4}). For more 
details see the caption of Fig. \ref{fig2}.} 
\label{fig5} 
\end{figure}

\begin{figure}[t]
\vspace{0.5cm} 
\begin{center}
\rotatebox{0}{\scalebox{0.4}{\includegraphics{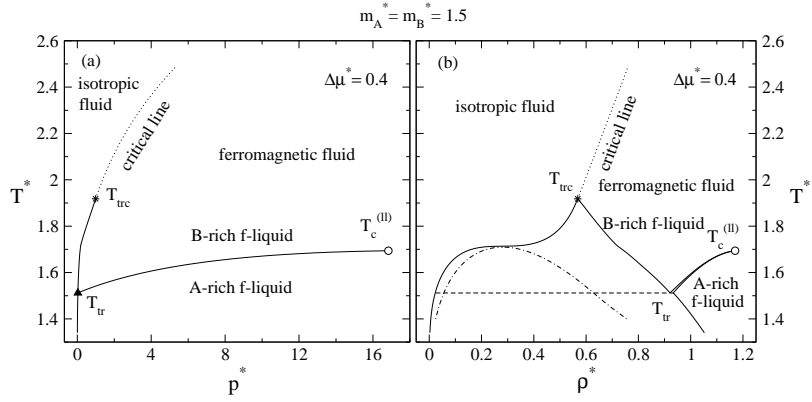}}}\\
\end{center}
\caption{Phase diagrams of a binary Stockmayer fluid mixture with
$m_A^{\star}=m_B^{\star}=1.5$ (compare Fig. \ref{fig1}(c)) in the
pressure-temperature (a) and density-temperature (b) plane for
$\Delta\mu^{\star}=0.4$; the LJ parameters are the same as in
Figs. \ref{fig2}-\ref{fig5}. For this choice of $\Delta\mu$, below the isotropic
vapor -- ferromagnetic A-rich liquid -- ferromagnetic B-rich liquid triple point
temperature $T_{tr}$ the isotropic vapor coexists with the ferromagnetic A-rich
liquid. Between $T_{tr}$ and $T_c^{(ll)}$ there are three possible phases: the
isotropic vapor, the ferromagnetic B-rich liquid, and the ferromagnetic A-rich
liquid. In (b) the two-phase region of A-rich and B-rich liquid coexistence
ending at $T_c^{(ll)}$ is rather narrow. Between $T_c^{(ll)}$ and the
tricritical temperature $T_{trc}$ the isotropic vapor coexists with the B-rich
ferromagnetic liquid. Above the tricritical temperature $T_{trc}$ the
transitions between the isotropic and ferromagnetic fluid are continuous and their
loci are given by the dotted critical line. The dash-dotted line in (b)
indicates the thermodynamically unstable isotropic liquid -- vapor coexistence.}  
\label{fig6}
\end{figure}

\begin{figure}[t]
\vspace{0.5cm} 
\begin{center}
\rotatebox{0}{\scalebox{0.4}{\includegraphics{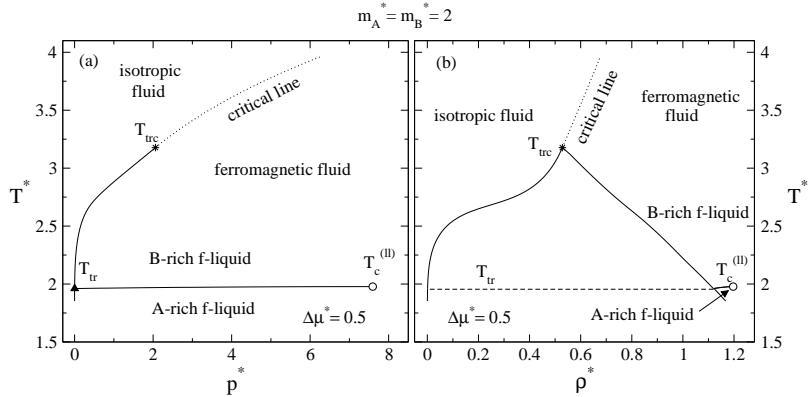}}}\\
\end{center}
\caption{Phase diagrams of a binary Stockmayer fluid mixture with
$m_A^{\star}=m_B^{\star}=2$ (compare Fig. \ref{fig1}(d)) in the
pressure-temperature (a) and density-temperature (b) plane for
$\Delta\mu^{\star}=0.5$; the LJ parameters are the same as in
Figs. \ref{fig2}-\ref{fig6}. Below the triple temperature $T_{tr}$ the
isotropic vapor coexists with a ferromagnetic A-rich liquid. The vapor phase
occurs at such small values of $p$ and $\rho$ that it is not visible on the
scales shown in (a) and (b). Within the very narrow temperature range
$T_{tr}<T<T_c^{(ll)}$ there are three possible phases: the isotropic vapor, the
ferromagnetic B-rich liquid, and the ferromagnetic A-rich liquid. Above the
ferromagnetic B-rich liquid -- ferromagnetic A-rich liquid critical temperature
$T_c^{(ll)}$ there is only coexistence between the isotropic fluid and the
B-rich ferromagnetic liquid. Above the tricritical temperature $T_{trc}$ these
phase transition are continuous forming the dotted critical line.}
\label{fig7}
\end{figure}

Figure \ref{fig1}(b) shows the schematic phase diagram of the binary liquid
mixture with the same LJ parameters but with $m_A^{\star}=m_B^{\star}=1$. The
occurrence of a ferromagnetic phase gives rise to the appearance of the sheet
$S_3$ which separates the isotropic liquid at high temperatures from the
ferromagnetic liquid at low temperatures. The nature of this sheet $S_3$
differs from that of $S_1$ and $S_2$ in that $S_3$ is the locus of
second-order phase transitions. In the thermodynamic space ($T,p,\Delta\mu$)
the surface $S_3$ is given by the simultaneous solution of Eqs. (\ref{sec2}) 
and (\ref{kont}). (With the knowledge of $\rho_A$, $\rho_B$, and $T$ 
Eq. (\ref{press}) gives the corresponding pressure $p$, using the fact that on $S_3$
$\alpha_i^{(0)}=1/2$ and $\alpha_i^{(l)}=0$ for $l\geq1$.) The intersection of
$S_3$ and $S_1$ is a line $M_1$ of critical end points. Along this line the
ferromagnetically critical liquid phase coexists with the isotropic vapor 
phase. The other line $M_2$ of critical end points is given by the
intersection of the sheets $S_2$ and $S_3$. The sheet $S_3$ divides the triple
line into two parts. The upper part describes isotropic vapor -- isotropic B-rich
liquid -- isotropic A-rich liquid three-phase coexistence, while the lower part
describes an isotropic vapor -- ferromagnetic B-rich liquid -- ferromagnetic
A-rich liquid coexistence. Both critical lines $L_1$ and $L_2$ are in the
isotropic part of the thermodynamic space ($T,p,\Delta\mu$). Figure \ref{fig3}
shows the corresponding results of our numerical calculations for a binary
Stockmayer fluid mixture with dipole moments $m_A^{\star}=m_B^{\star}=1$ for
$\Delta\mu^{\star}=0.3$. The intersection of the plane $\Delta\mu^{\star}=0.3$
with the sheet $S_3$ yields the critical (dotted) line in
Fig. \ref{fig3}(a). Its intersection with the first-order coexistence curve
between vapor and ferromagnetic A-rich liquid is the critical end point CEP
with $T_{CEP}^{\star}=1.082$, $p_{CEP}^{\star}=0.03$, and
$\rho_{CEP}^{\star}=0.74$ at $\Delta\mu^{\star}=0.3$. Below $T_{CEP}$ the
isotropic vapor phase coexists with the ferromagnetic A-rich liquid. The
critical line of second-order phase transitions divides the A-rich liquid into
an isotropic A-rich liquid phase and into a ferromagnetic A-rich liquid phase. 
In the temperature range $T_{CEP}<T<T_{tr}$ with $T_{tr}^{\star}=1.12$ the
isotropic vapor coexists with an isotropic A-rich liquid phase. Above the triple
point temperature there is also coexistence between the isotropic B-rich
liquid and the isotropic A-rich liquid. This first-order phase transition ends
in the critical temperature $T_c^{(ll)\star}=1.404$. The critical temperature
of the isotropic vapor -- isotropic B-rich liquid coexistence is
$T_c^{(lv)\star}=1.296$.

Upon increasing the dipole moments the phase diagram shown in
Fig. \ref{fig1}(b) changes in two respects. First, the sheet $S_3$ is raised
due to the strengthening of orientational ordering. Second, close to the line
$M_1$ the character of the ferromagnetic - isotropic phase transition on $S_3$
becomes first-order such that a line of tricritical points ($TC$) emerges 
from $M_1$ 
and separates the sheet $S_3$ into a continuous part (back part at high $p$
and hatched) and a first-order part (front part at lower $p$ and unhatched). 
This implies that $M_1$ turns into a triple line 
$TL^{\star}$. Upon further increase of $m_A$ and $m_B$ this sheet $S_3$,
divided by the $TC$-line, raises finally above the line $L_2$. Moreover the
upper part of $S_1$, i.e., above $TL^{\star}$ disintegrates into a left and
right part leaving a gap between them. This means that $TL^{\star}$ is split
into $TL_1$ and $TL_1^{\prime}$ and $L_1$ into $L_1$ and $L_1^{\prime}$ in
Fig. \ref{fig1}(c). This also generates two critical end points $CEP_1$ and
$CEP_1^{\prime}$ which are raised above $CEP$. Figure \ref{fig1}(c),
corresponding to $m_A^{\star}=m_B^{\star}=1.5$, displays this latter structure
with $S_1\cup{S_3}$ denoting the sheet which emerged from the lower part of
$S_1$ in Fig. \ref{fig1}(b) and $S_3$ whereas the winglike structures in
Fig. \ref{fig1}(c) are the remnants of the upper part of $S_1$ in
Fig. \ref{fig1}(b). They are connected to the sheet $S_1\cup{S_3}$ along the
triple lines $TL_1$ and $TL_1^{\prime}$ and bounded by the liquid-vapor
critical lines $L_1$ and $L_1^{\prime}$. The isotropic vapor -- isotropic fluid
-- ferromagnetic liquid triple lines $TL_1$ and $TL_1^{\prime}$ end in critical
end points $CEP_1$ and $CEP_1^{\prime}$. $S_2$ and $S_1\cup{S_3}$ intersect
along the triple line $TL$ which ends at the critical end point $CEP$.

Figure \ref{fig4} shows the corresponding results for the projections of the
triple lines and of the liquid-vapor critical lines of a binary Stockmayer
fluid mixture with dipole moments $m_A^{\star}=m_B^{\star}=1.5$ onto a plane
$p=const$. The triple line $TL$ describes the isotropic vapor -- ferromagnetic
A-rich liquid -- ferromagnetic B-rich liquid three-phase coexistence with the
critical end point temperature $T_{CEP}^{\star}=1.614$. The triple lines $TL_1$
and $TL_1^{\prime}$ associated with the two winglike surfaces (see
Fig. \ref{fig1}(c)) describe the isotropic vapor -- isotropic fluid --
ferromagnetic A-rich liquid and isotropic vapor -- isotropic fluid --
ferromagnetic B-rich liquid three-phase coexistences with the critical end
point temperatures $T^{\star}_{CEP_1^{\prime}}=1.764$ and
$T^{\star}_{CEP_1}=1.747$.

Figure \ref{fig5} presents our numerical results for a binary Stockmayer fluid
mixture with dipole moments $m_A^{\star}=m_B^{\star}=1.5$ for
$\Delta\mu^{\star}=4$. For this choice of $\Delta\mu$ at the triple point $T_{tr}$ 
the following three phases coexist: vapor, isotropic fluid, and 
ferromagnetic liquid. At low temperatures $T<T_{tr}$ with
$T_{tr}^{\star}=1.768$ the ferromagnetic liquid coexists with the isotropic
vapor. Above the triple point temperature, two isotropic fluids coexist at low
and moderate densities, becoming identical above the liquid-vapor critical
temperature $T_c^{(lv)\star}=1.788$. (The corresponding critical pressure and
density are $p_c^{(lv)\star}=0.185$ and $\rho_c^{(lv)\star}=0.286$,
respectively.) At higher densities, the isotropic liquid and magnetic liquid
are separated by a first-order phase transition which becomes second order
above the tricritical temperature $T_{trc}^{\star}=1.954$. (The corresponding
tricritical pressure and density are $p_{trc}^{\star}=0.964$ and
$\rho_{trc}^{\star}=0.579$, respectively.) The critical line of second-order
phase transitions divides the liquid range into an isotropic liquid phase and
into a ferromagnetic liquid phase. At this value of the chemical potential
difference there is no A-rich liquid -- B-rich liquid demixing phase
separation.

Figure \ref{fig6} shows our numerical results for a binary Stockmayer fluid
mixture with dipole moments $m_A^{\star}=m_B^{\star}=1.5$ for
$\Delta\mu^{\star}=0.4$. At the triple point temperature $T_{tr}$ the coexisting 
phases are: vapor, 
ferromagnetic B-rich liquid, and ferromagnetic A-rich liquid. At low
temperatures $T<T_{tr}$ with $T_{tr}^{\star}=1.512$ the isotropic vapor phase
coexists with the ferromagnetic A-rich liquid. Above the triple point
temperature, an isotropic vapor coexists with a ferromagnetic B-rich liquid and
a ferromagnetic B-rich liquid coexists with a ferromagnetic A-rich liquid. The
two ferromagnetic liquid phases become identical beyond the liquid-liquid
critical point $T_c^{(ll)}$ with $T_c^{(ll)\star}=1.694$. (The corresponding 
critical pressure and density are $p_c^{(ll)\star}=16.83$ and 
$\rho_c^{(ll)\star}=1.17$, respectively, at $\Delta\mu^{\star}=0.4$.) Above
the ferromagnetic A-rich liquid -- ferromagnetic B-rich liquid critical
temperature $T_c^{(ll)}$ the isotropic liquid and ferromagnetic B-rich liquid are
separated by a first-order transition which turns second order above the
tricritical temperature $T_{trc}^{\star}=1.918$. (The corresponding
tricritical pressure and density are $p_{trc}^{\star}=0.994$ and
$\rho_{trc}^{\star}=0.568$, respectively, at $\Delta\mu^{\star}=0.4$.) At
temperatures $T>T_{trc}$ one finds second-order ferromagnetic phase transitions 
as given by the critical line.

Fig. \ref{fig1}(d) shows that by further increasing the dipole moments
($m_A^{\star}=m_B^{\star}=2$) the winglike surfaces associated with the
isotropic liquid -- isotropic vapor transition disappear and only the
two sheets $S_2$ and $S_1\cup{S_3}$ remain. Accordingly there is only one
triple line $TL$, which describes the isotropic vapor, ferromagnetic A-rich
liquid, and ferromagnetic B-rich liquid three-phase coexistence and which ends
at the critical end point $CEP$ where $L_2$ hits $S_1\cup{S_3}$. Since the
sheet $S_2$ bounded by $L_2$ lies beneath the sheet $S_1\cup{S_3}$ all
first-order liquid-liquid phase separations involve ferromagnetic
phases.

Figure \ref{fig7} shows our numerical results for a binary Stockmayer
fluid mixture with dipole moments $m_A^{\star}=m_B^{\star}=2$ for
$\Delta\mu^{\star}=0.5$. The topology of Fig. \ref{fig7} is the same as in
Fig. \ref{fig6}, only the numerical values for the triple, critical, and
tricritical points are different. Below the triple point temperature
$T_{tr}^{\star}=1.961$ (with $p_{tr}^{\star}=0.002$ at $\Delta\mu^{\star}=0.5$)
an isotropic vapor coexists with a ferromagnetic A-rich liquid. The
ferromagnetic liquid-liquid critical point is given by $T_c^{(ll)\star}=1.98$,
$p_c^{(ll)\star}=7.59$, and $\rho_c^{(ll)\star}=1.2$, while the tricritical
point is given by $T_{trc}^{\star}=3.17$, $p_{trc}^{\star}=2.06$, and
$\rho_{trc}^{\star}=0.528$.

\subsection{Binary liquid mixtures of non-polar and polar particles of equal size}

\begin{figure}[t]
\vspace{0.5cm} 
\begin{center}
\rotatebox{0}{\scalebox{0.45}{\includegraphics{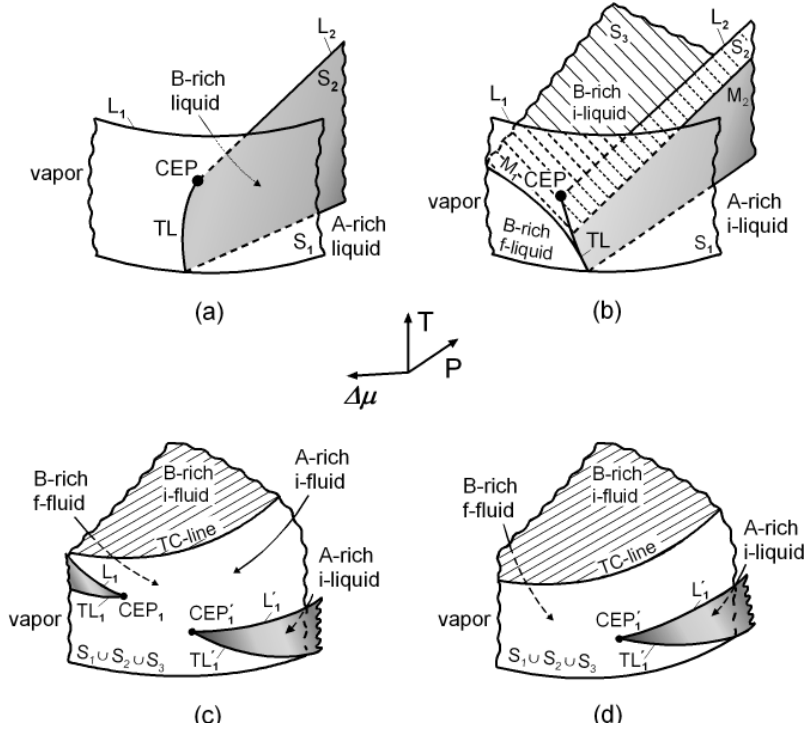}}}\\
\end{center}
\caption{Schematic bulk phase diagrams of binary Stockmayer fluid
mixtures in the thermodynamic space of temperature $T$, pressure $p$,
and chemical potential difference $\Delta\mu=\mu_B-\mu_A$ of
the non-polar ($m_A^{\star}=0$) and polar ($m_B^{\star}{\neq}0$) components. 
For reference purposes the phase diagram in (a) 
represents the same non-polar ($m_A^{\star}=m_B^{\star}=0$) binary 
fluid mixture as in Fig.\ref{fig1}(a). The phase diagram in (b) 
represents a non-polar -- dipolar ($m_A^{\star}=0$, $m_B^{\star}=1$) 
binary fluid mixture for which in comparison with (a) a third sheet 
$S_3$ (hatched) appears, 
which separates the B-rich isotropic liquid from the B-rich ferromagnetic 
liquid. $S_3$ is a sheet of second-order phase transitions. 
$M_1$ and $M_2$ represent lines of critical end points at the intersection 
of $S_3$ with $S_1$ and $S_2$, respectively. In (a)-(d) the A-rich 
liquid is always isotropic.  Upon increasing the dipole 
moment of the component $B$ ($m_A^{\star}=0$, 
$m_B^{\star}=1.5$) in (c) the tilted sheet $S_3$ has moved above
$L_1$ and $L_2$ and has merged with $S_1$ and $S_2$ into the sheet
$S_1{\cup}S_2{\cup}S_3$ with a tricritical line ($TC$-line), which
separates first-order transitions (lower part of the sheet) from
second-order transitions (hatched upper part). The loci of isotropic
liquid -- vapor transitions are reduced to two winglike surfaces, which
are connected to the surface $S_1{\cup}S_2{\cup}S_3$ along the triple
lines $TL_1$ and $TL_1^{\prime}$. These triple lines end in critical
end points $CEP_1$ and $CEP_1^{\prime}$. The liquid-vapor critical
lines $L_1$ and $L_1^{\prime}$ are the upper edges of the wings. Upon
a further increase of the dipole moment of component $B$
($m_A^{\star}=0$, $m_B^{\star}=2$) one obtains the phase diagram shown
in (d) where there is only one sheet $S_1{\cup}S_2{\cup}S_3$ and one
wing attached to it. The surface $S_1{\cup}S_2{\cup}S_3$ denotes the 
isotropic vapor -- ferromagnetic liquid transition, 
while the wing is associated with the A-rich isotropic liquid -- isotropic 
vapor transition.}        
\label{fig8}
\end{figure}

In this subsection we present our results for phase diagrams 
of non-polar ($m_A^{\star}=0$) -- polar ($m_B^{\star}\neq0$) binary mixtures. 
The corresponding schematic phase diagrams are shown in Fig. \ref{fig8}. 
In order to provide a convenient visual comparison 
Fig. \ref{fig8}(a) shows again as a reference case the schematic phase diagram of a 
non-polar 
binary Lennard-Jones fluid mixture ($m_A^{\star}=m_B^{\star}=0$) with the 
parameter set $\epsilon_{BB}/\epsilon_{AA}=0.95$, $\epsilon_{AB}/\epsilon_{AA}
=0.75$ and $\sigma_{AA}=\sigma_{BB}=\sigma_{AB}$. Figure \ref{fig8}(b) shows 
the schematic phase diagram of the binary liquid mixture with the same LJ 
parameters but with $m_A^{\star}=0$ and $m_B^{\star}=1$. Similar to 
Fig. \ref{fig1}(b) the occurrence of a ferromagnetic phase gives rise to the 
appearance of the sheet $S_3$ of second-order phase transitions 
which separates the isotropic liquid phase from 
the ferromagnetic liquid phase. 
The ferromagnetic phase appears only in the B-rich liquid phase and 
at low temperatures 
while the A-rich liquid phase remains isotropic even at low temperatures. 
This behavior reflects the polar ($B$) and non-polar ($A$) character of molecules. 
In accordance with this asymmetry the sheet $S_3$ is tilted, ends at $S_2$ and 
thus does not cut through the sheet $S_2$. This leads to the line $M_2$ of 
critical end points. The intersection of $S_3$ and $S_1$ forms another line $M_1$ 
of critical end points. The triple line ($TL$) is given 
by the intersection of $S_1$ and $S_2$. The line $M_1$ of critical end points 
divides the triple line into two parts. 
The upper part of the triple line describes isotropic vapor -- 
isotropic B-rich liquid -- isotropic A-rich liquid three-phase coexistence as 
in the case of symmetric dipolar mixtures with dipole moments 
$m_A^{\star}=m_B^{\star}=1$ (see Fig. \ref{fig1}(b)). The lower part
of the triple line describes isotropic vapor -- ferromagnetic B-rich 
liquid -- isotropic A-rich liquid coexistence.


\begin{figure}[t]
\vspace{0.5cm} 
\begin{center}
\rotatebox{0}{\scalebox{0.4}{\includegraphics{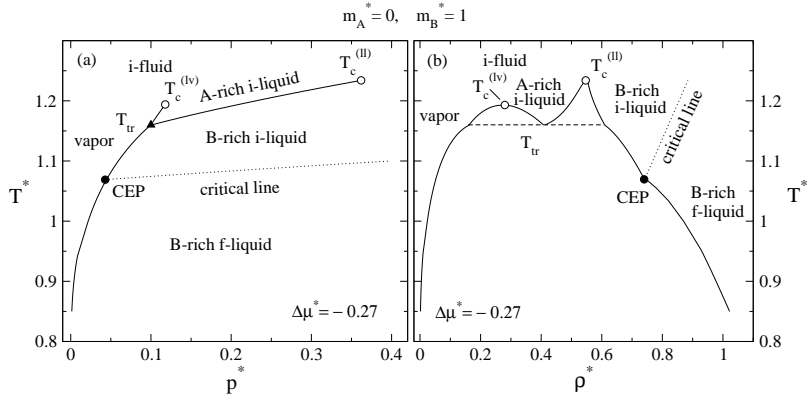}}}\\
\end{center}
\caption{Phase diagrams of a non-polar -- polar ($m_A^{\star}=0$, 
$m_B^{\star}=1$) binary Stockmayer fluid mixture (compare Fig. \ref{fig8}(b)) 
in the pressure-temperature (a) and density-temperature (b) plane for 
$\Delta\mu^{\star}=-0.27$. The LJ parameters are the same as in 
Fig. \ref{fig2}. Below the critical end point ($CEP$) temperature 
$T_{CEP}$ the isotropic vapor coexists with the B-rich ferromagnetic 
liquid. Between $T_{CEP}$ and $T_{tr}$ the isotropic vapor coexists 
with the B-rich isotropic liquid. Between $T_{tr}$ and $T_c^{(lv)}$
there are three possible phases: the vapor, the A-rich isotropic liquid,
and the B-rich isotropic liquid. Between $T_c^{(lv)}$ and $T_c^{(ll)}$ 
only A-rich isotropic liquid -- B-rich isotropic liquid coexistence is possible. 
The dotted line denotes a line of second-order ferromagnetic phase transitions. It 
ends at a critical end point $CEP$.}
\label{fig9} 
\end{figure}

\begin{figure}[t] 
\vspace{0.5cm}
\begin{center}
\rotatebox{0}{\scalebox{0.4}{\includegraphics{fig10.eps}}}\\
\end{center}
\caption{Phase diagrams of a binary Stockmayer fluid mixture for
$m_A^{\star}=0$ and $m_B^{\star}=1$ (compare Fig. \ref{fig8}(b)) in
the pressure-temperature (a) and density-temperature (b) plane for
$\Delta\mu^{\star}=-0.54$; the LJ parameters are the same as in
Fig. \ref{fig2}. Below the triple point temperature $T_{tr}$ the
isotropic vapor coexists with the B-rich ferromagnetic liquid. Between
$T_{tr}$ and $T_c^{(lv)}$ there are three possible phases: the vapor,
the A-rich isotropic liquid, and the B-rich ferromagnetic liquid. The
A-rich isotropic liquid -- B-rich ferromagnetic liquid coexistence does
not end in a critical point within the range of fluid densities.}
\label{fig10}
\end{figure}


\begin{figure}[t] 
\vspace{0.5cm}
\begin{center}
\rotatebox{0}{\scalebox{0.4}{\includegraphics{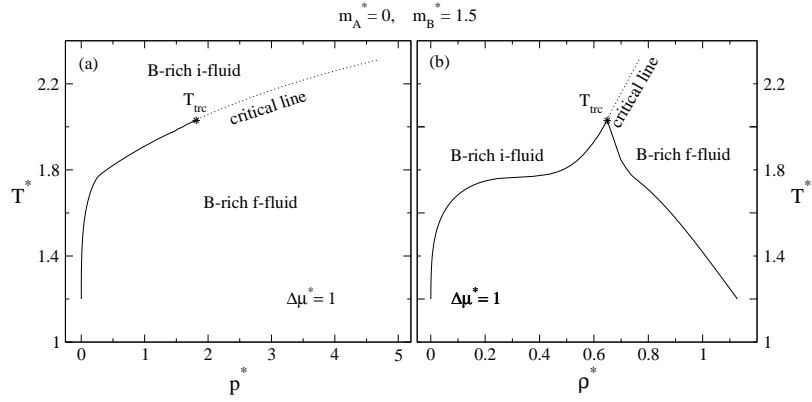}}}\\
\end{center}
\caption{Phase diagrams of a binary Stockmayer fluid mixture for 
$m_A^{\star}=0$ and $m_B^{\star}=1.5$ (compare Fig. \ref{fig8}(c)) 
in the pressure-temperature (a) and density-temperature (b) plane for 
$\Delta\mu^{\star}=1$; the LJ parameters are the same as in 
Fig. \ref{fig2}. Below the tricritical temperature $T_{trc}$ the 
B-rich isotropic fluid coexists with the B-rich ferromagnetic 
fluid. This first-order phase transition turns into a second-order 
phase transition at the tricritical temperature $T_{trc}$. Above $T_{trc}$ 
the dotted line represents the line of second-order phase transitions.}
\label{fig11} 
\end{figure}


\begin{figure}[t] 
\vspace{0.5cm}
\begin{center}
\rotatebox{0}{\scalebox{0.4}{\includegraphics{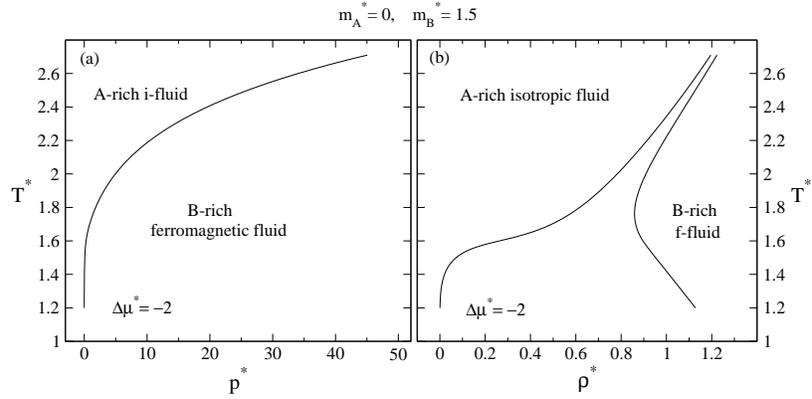}}}\\
\end{center}
\caption{Phase diagrams of a binary Stockmayer fluid mixture for 
$m_A^{\star}=0$ and $m_B^{\star}=1.5$ (compare Fig. \ref{fig8}(c))
in the pressure-temperature (a) and density-temperature (b) plane for
$\Delta\mu^{\star}=-2$; the LJ parameters are the same as in
Fig. \ref{fig2}. At this value of the chemical potential difference
only the A-rich isotropic liquid can coexist along a line of 
first-order phase transitions with the B-rich 
ferromagnetic liquid. The liquid-liquid coexistence does not end in a 
critical point within the range of physically relevant fluid densities.}   
\label{fig12} 
\end{figure}


\begin{figure}[t]
\vspace{0.5cm} 
\begin{center}
\rotatebox{0}{\scalebox{0.4}{\includegraphics{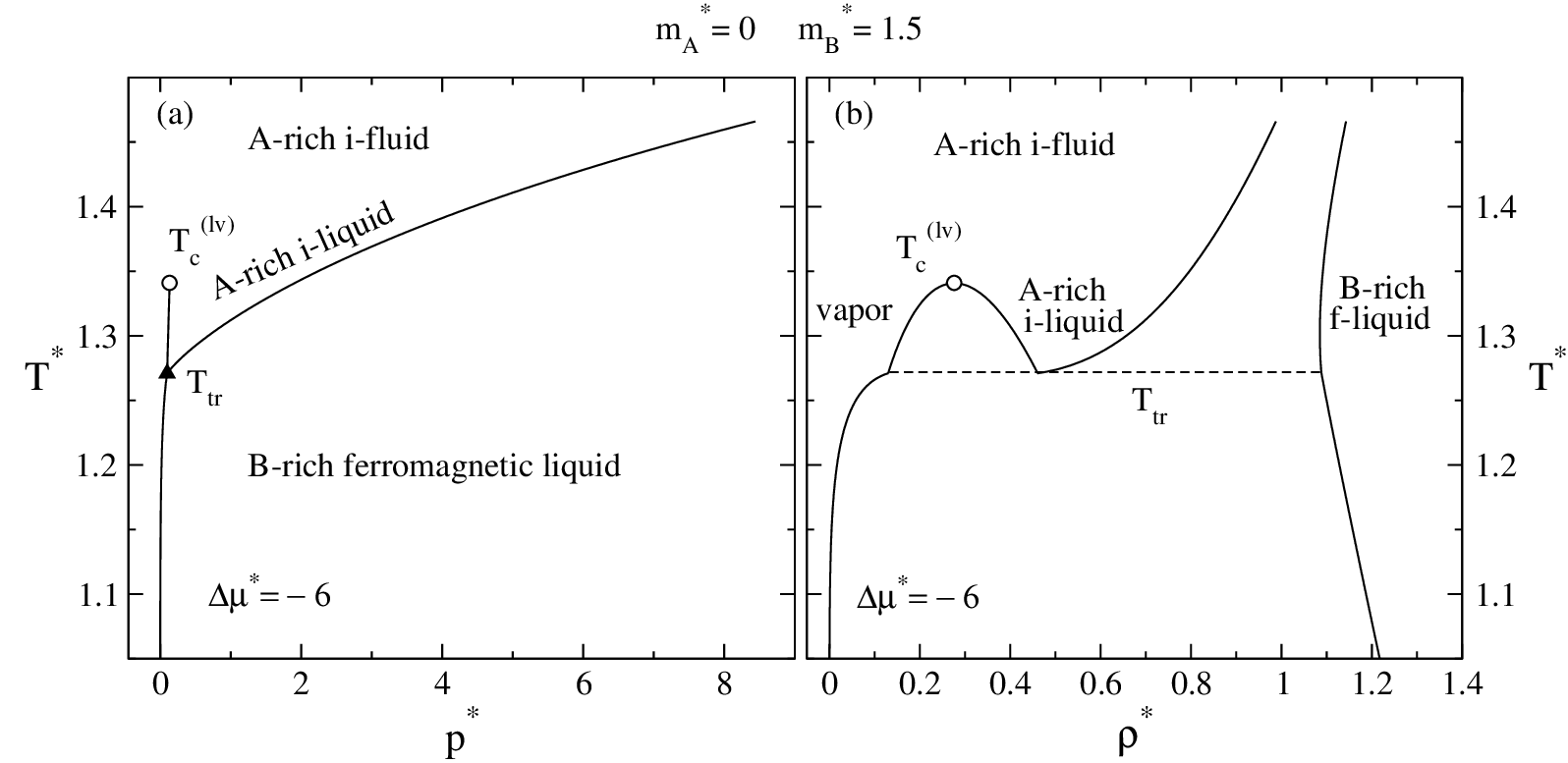}}}\\
\end{center}
\caption{Phase diagrams of a binary Stockmayer fluid mixture for 
$m_A^{\star}=0$ and $m_B^{\star}=1.5$ (compare Fig. \ref{fig8}(c)) 
in the pressure-temperature (a) and density-temperature (b) plane for 
$\Delta\mu^{\star}=-6$; the LJ parameters are the same as in 
Fig. \ref{fig2}. Below the triple point temperature $T_{tr}$ the 
isotropic vapor coexists with the B-rich ferromagnetic liquid. Between 
$T_{tr}$ and $T_c^{(lv)}$ there are three possible phases: the vapor, 
the A-rich isotropic liquid, and the B-rich ferromagnetic liquid. The 
A-rich isotropic liquid -- B-rich ferromagnetic liquid coexistence 
does not end in a critical point within the range of physically 
relevant fluid densities.}
\label{fig13}
\end{figure}

Figure \ref{fig9} shows our numerical findings for a binary Stockmayer fluid 
mixture with dipole moments $m_A^{\star}=0$ and $m_B^{\star}=1$ for 
$\Delta\mu^{\star}=-0.27$. Figure \ref{fig9}(a) demonstrates that for this
value of the chemical potential difference the plane
$\Delta\mu^{\star}=-0.27$ cuts through the upper part of the triple
line $TL$ in Fig. \ref{fig8}(b) so that the corresponding triple point $T_{tr}$ 
describes an isotropic vapor -- A-rich isotropic liquid -- B-rich isotropic 
liquid coexistence with $T_{tr}^{\star}=1.16$ and $p_{tr}^{\star}=0.101$. 
The intersection of this plane with the sheet $S_3$ yields the dotted critical line. 
The intersection of the plane $\Delta\mu^{\star}=-0.27$ with the line $M_1$ 
in Fig. \ref{fig8}(b) gives the critical end point $CEP$ where an 
isotropic vapor coexists with a ferromagnetically critical B-rich liquid. 
The thermodynamic parameters of this critical end point $CEP$ are
given by $T_{CEP}^{\star}=1.07$, $p_{CEP}^{\star}=0.043$, and 
$\rho_{CEP}^{\star}=0.74$, respectively, for $\Delta\mu^{\star}=-0.27$. 
Below the temperature $T_{CEP}$  the isotropic vapor phase coexists with 
the ferromagnetic B-rich liquid phase. Between the temperatures 
$T_{CEP}$ and $T_{tr}$ two phases coexist: the isotropic vapor and the 
B-rich isotropic liquid. Above $T_{tr}$ up to the liquid-vapor critical temperature 
($T_c^{(lv)\star}=1.19$) coexistence is possible between the vapor and the 
A-rich isotropic liquid phase and between the A-rich isotropic liquid 
and the B-rich isotropic liquid. In the temperature range 
$T_c^{(lv)}<T<T_c^{(ll)}$ with $T_c^{(ll)\star}=1.23$ coexistence is possible only 
between the isotropic A-rich and B-rich liquid phases. 
(The corresponding critical densities are $\rho_c^{(lv)\star}=0.31$ 
and $\rho_c^{(ll)\star}=0.55$, respectively.) 

Figure \ref{fig10} shows our numerical results for a binary
Stockmayer fluid mixture with dipole moments $m_A^{\star}=0$ and
$m_B^{\star}=1$ for $\Delta\mu^{\star}=-0.54$. Figure \ref{fig10}(a)
displays that for this value of the chemical potential difference the
plane $\Delta\mu^{\star}=-0.54$ cuts through the lower part of the
triple line $TL$ in Fig. \ref{fig8}(b) so that the corresponding 
triple point $T_{tr}$ describes an isotropic vapor -- A-rich isotropic 
liquid -- B-rich ferromagnetic liquid coexistence with $T_{tr}^{\star}=1.05$ and 
$p_{tr}^{\star}=0.045$. Below the triple temperature $T_{tr}$ the vapor phase 
coexists with the B-rich ferromagnetic liquid phase. Above $T_{tr}$ up
to the liquid-vapor critical temperature ($T_c^{(lv)\star}=1.21$)
coexistence is possible between the vapor and the A-rich isotropic liquid phase 
and between the A-rich isotropic liquid and the B-rich ferromagnetic
liquid. The critical pressure and density of liquid-vapor
coexistence are $p_c^{(lv)\star}=0.122$ and
$\rho_c^{(lv)\star}=0.285$, respectively. 
The A-rich isotropic liquid -- B-rich ferromagnetic liquid phase coexistence 
extends to higher temperatures. The corresponding liquid-liquid critical point 
is not attainable at physically relevant fluid densities. 
(At high densities the system freezes, but the corresponding solid phases are not 
captured within the framework of the present theory.)

Upon increasing the dipole moment of component $B$ the phase diagram 
shown in Fig. \ref{fig8}(b) changes considerably. First, 
the sheet $S_3$ moves up due to the strengthening of the dipolar 
interaction. Like in the case of symmetrical dipolar mixtures 
($m_A^{\star}=m_B^{\star}=1.5$, see Fig. \ref{fig1}(c)) close 
to the line $M_1$ the 
ferromagnetic-isotropic phase transition on the sheet $S_3$ becomes 
first order. This means that a line $TC$ of tricritical points emerges from
$M_1$ and separates the sheet $S_3$ into a continuous part and a first-order 
part. In turn this implies that $M_1$ turns into a triple line 
$TL^{\star}$. In the next step due to the rise of the sheet $S_3$ (divided by the 
$TC$-line) the line $M_2$ rises, too, and merges with the line $L_2$ so that
$S_3$ and $S_2$ form a single sheet $S_2{\cup}S_3$. 
Upon a further increase of $m_B$ the upper part of $S_1$ 
disintegrates into a left and a right part leaving a gap between them, 
where $S_1$ merges with $S_2{\cup}S_3$ forming the surface 
$S_1{\cup}S_2{\cup}S_3$. The remnants of the upper part of the 
sheet $S_1$ appear as asymmetrically winglike structures, where the 
triple lines $TL_1$ and $TL_1^{\prime}$ stem from the triple line 
$TL^{\star}$ which has split while the split line $L_1$ generates 
the liquid-vapor critical lines $L_1$ and 
$L_1^{\prime}$ (see Fig. \ref{fig8}(c)). The intersections of the 
corresponding triple and critical lines give two critical end points 
$CEP_1$ and $CEP_1^{\prime}$. Figure \ref{fig8}(c) corresponds to 
$m_A^{\star}=0$ and $m_B^{\star}=1.5$ and displays this final structure 
with $S_1{\cup}S_2{\cup}S_3$ denoting the surface which has emerged from 
the sheets $S_1$, $S_2$, and $S_3$ (shown in Fig. \ref{fig8}(b)) upon the 
increase of the dipole moment $m_B$. 

Figure \ref{fig11} shows our numerical results for a binary
Stockmayer fluid mixture with dipole moments $m_A^{\star}=0$ and
$m_B^{\star}=1.5$ for $\Delta\mu^{\star}=1$. For this value of 
$\Delta\mu^{\star}$ the figure displays a vertical intersection of the phase 
diagram shown in Fig. \ref{fig8}(c) between the two critical end points 
$CEP_1$ and  $CEP_1^{\prime}$ and close to $CEP_1$.
At low temperatures the B-rich isotropic fluid
and the B-rich ferromagnetic fluid are separated by a first-order 
phase transition which becomes second order above the tricritical 
temperature $T_{trc}^{\star}=2.03$. (The tricritical pressure and 
density are $p_{trc}^{\star}=1.82$ and $\rho_{trc}^{\star}=0.65$, respectively.) 
The dotted critical line corresponds to a cut through the second-order part of the 
surface $S_1{\cup}S_2{\cup}S_3$. 

Figure \ref{fig12} shows our numerical findings for a binary
Stockmayer fluid mixture with dipole moments $m_A^{\star}=0$ and 
$m_B^{\star}=1.5$  for $\Delta\mu^{\star}=-2$. This figure displays
a vertical cut through the phase diagram in Fig. \ref{fig8}(c) between 
$CEP_1$ and  $CEP_1^{\prime}$ but close to the critical end point $CEP_1^{\prime}$. 
For this value of the chemical potential difference the A-rich isotropic 
fluid can coexist with the B-rich ferromagnetic fluid. Note that the B-rich isotropic 
fluid and the A-rich isotropic fluid are not separated by a demixing transition; 
in Fig. \ref{fig8}(c) both are in front of $S_1{\cup}S_2{\cup}S_3$. 
The liquid-liquid critical point lies outside the physically 
relevant fluid densities.

Figure \ref{fig13} shows our numerical results for a binary
Stockmayer fluid mixture with dipole moments $m_A^{\star}=0$ and 
$m_B^{\star}=1.5$ for $\Delta\mu^{\star}=-6$. In this case the plane 
$\Delta\mu^{\star}=-6$ cuts through the wing on the right side of the surface 
$S_1{\cup}S_2{\cup}S_3$ in  Fig. \ref{fig8}(c). For this choice of
$\Delta\mu$ at the triple point $T_{tr}$ the following three phases coexist: 
vapor, A-rich isotropic liquid, and B-rich ferromagnetic 
liquid. At low temperatures $T<T_{tr}$ with $T_{tr}^{\star}=1.27$ 
($p_{tr}^{\star}=0.1$) the B-rich ferromagnetic liquid coexists with 
the isotropic vapor. Above the triple point temperature at low densities the 
vapor coexists  with the A-rich isotropic liquid, whereas at high densities 
the A-rich isotropic liquid coexists with the B-rich ferromagnetic liquid. 
The vapor and the A-rich isotropic liquid merge at the liquid-vapor critical point 
$T_c^{(lv)\star}=1.34$,  $p_c^{\star}=0.13$, and $\rho_c^{\star}=0.28$. 
The A-rich isotropic liquid -- B-rich ferromagnetic liquid critical point 
occurs outside the physically relevant fluid densities.

Figure \ref{fig8}(d) shows that by further increasing the dipole 
moment of the component $B$ ($m_A^{\star}=0$, $m_B^{\star}=2$) the 
winglike structure on the left hand side in  Fig. \ref{fig8}(c), which is 
associated with the B-rich isotropic liquid -- vapor transitions, disappears 
so that there remain only the two sheets $S_1{\cup}S_2{\cup}S_3$ and the wing 
on the right hand side associated with the A-rich isotropic liquid -- vapor transitions. 
Accordingly there is only one triple line $TL_1^{\prime}$, 
which describes the isotropic vapor -- A-rich isotropic liquid -- B-rich 
ferromagnetic liquid three-phase coexistence. It ends at 
the critical end point $CEP_1^{\prime}$ where $L_1^{\prime}$ hits the surface 
$S_1{\cup}S_2{\cup}S_3$. Since the topology of the phase diagram in 
Fig. \ref{fig8}(d) is the same as that of Fig. \ref{fig8}(c) (only the 
left winglike surface in Fig. \ref{fig8}(c) is missing), the topologies of 
different cuts at given chemical potentials are in agreement  with the 
topologies of the phase diagrams in Figs. \ref{fig11}-\ref{fig13}. 
Therefore we refrain from presenting our detailed numerical results for 
Stockmayer mixtures with dipole moments $m_A^{\star}=0$ and $m_B^{\star}=2$.

\subsection{Binary liquid mixtures of polar particles of different size}

In this subsection we present phase diagrams of binary Stockmayer liquid 
mixtures of polar particles ($m_A^{\star}{\neq}m_B^{\star}$) of different size. 
Since in the Stockmayer interaction potential the size parameters appear in the LJ 
contribution to the potential, we first study the effect of varying the ratio 
$\sigma_{BB}/\sigma_{AA}$ on the phase diagrams of binary LJ fluid mixtures 
($m_A^{\star}=m_B^{\star}=0$) while keeping the energy parameters of the LJ 
potential fixed.

\begin{figure}[t] 
\vspace{0.5cm}
\begin{center}
\rotatebox{0}{\scalebox{0.4}{\includegraphics{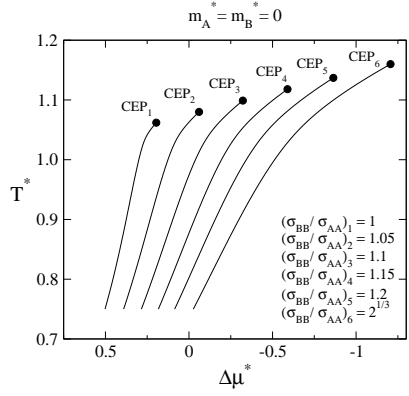}}}\\
\end{center}
\caption{Projections of the vapor -- A-rich isotropic liquid -- B-rich isotropic liquid 
triple line $TL$ (see Fig. \ref{fig1}) of binary Lennnard-Jones fluid mixtures 
($m_A^{\star}=m_B^{\star}=0$) onto a plane $p=const$ for 
$\epsilon_{BB}/\epsilon_{AA}=0.95$, $\epsilon_{AB}/\epsilon_{AA}=0.75$, and 
various size ratios $\sigma_{BB}/\sigma_{AA}$. The triple lines end at different critical 
end points ($CEP$s, see Fig. \ref{fig1}). 
Note that as in Figs. \ref{fig1} and \ref{fig4} $\Delta\mu$ increases towards the left.}
\label{fig14}
\end{figure}

Figure \ref{fig14} shows the projections of the triple lines of binary LJ fluid mixtures 
with different particle size ratios onto a plane $p=const$. These triple lines describe the 
vapor -- A-rich isotropic liquid -- B-rich isotropic liquid three-phase coexistence. 
The curves show that the sheet $S_2$, which separates the A-rich and B-rich liquid phases 
(see Fig. \ref{fig1}(a)) is increasingly  tilted upon increasing the size ratio 
$\sigma_{BB}/\sigma_{AA}$. The critical end point temperature increases and the chemical 
potential difference $\Delta\mu^{\star}_{CEP}$ at the end point turns more negative, i.e., 
it is shifted towards the region of fluid rich in the small particles. 
This means that for large size ratios three phase coexistence occurs for a wider range of 
thermodynamic parameters. 

\begin{figure}[t] 
\vspace{0.5cm}
\begin{center}
\rotatebox{0}{\scalebox{0.4}{\includegraphics{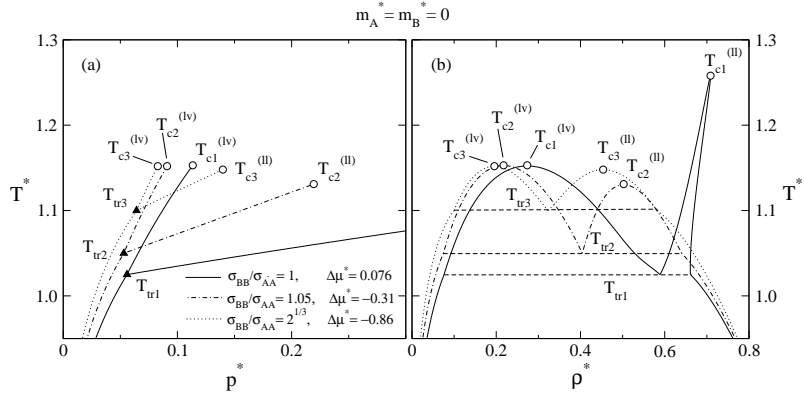}}}\\
\end{center}
\caption{Phase diagrams of the binary Lennard-Jones fluid mixtures in the 
pressure-temperature (a) and density temperature (b) plane for different size ratios 
$\sigma_{BB}/\sigma_{AA}$ and $\Delta\mu^{\star}$. The LJ energy parameters are 
the same as in Fig. \ref{fig14}. In (a) all liquid-vapor coexistence curves 
separate the vapor phase from the B-rich isotropic liquid phase. The liquid-liquid 
coexistence curves separate the B-rich isotropic liquid phase from the 
A-rich isotropic liquid phase. In (b) the full, dashed-dotted, and dotted curves 
enclose two-phase regions whereas the dashed lines indicate three-phase coexistence. 
The line code in (b) corresponds to that in (a) with the same choices of the size ratio 
$\sigma_{BB}/\sigma_{AA}$ and $\Delta\mu^{\star}$.}
\label{fig15}
\end{figure}

\begin{figure}[t] 
\vspace{0.5cm}
\begin{center}
\rotatebox{0}{\scalebox{0.45}{\includegraphics{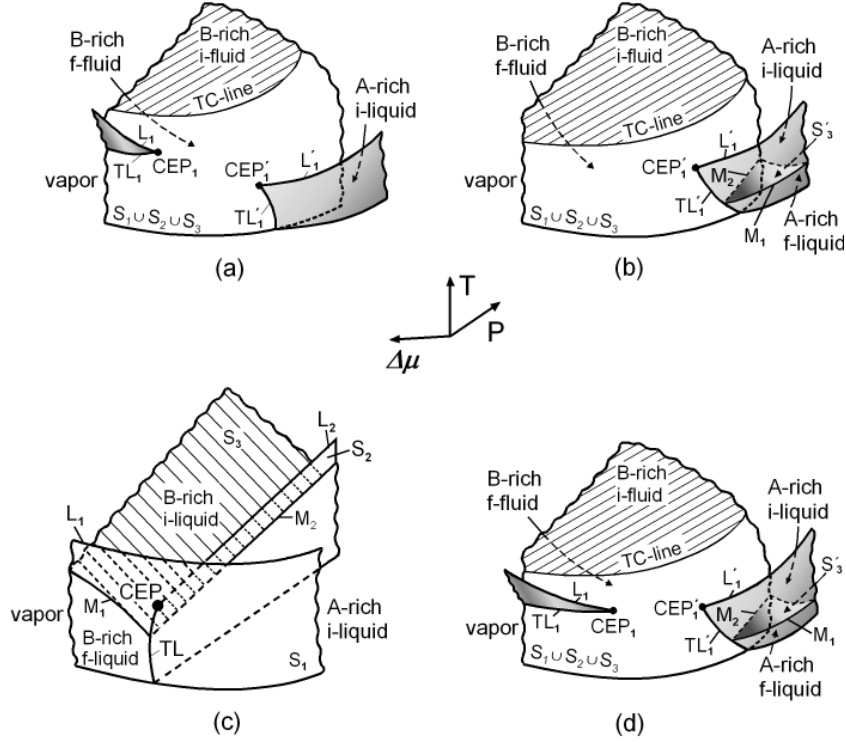}}}\\
\end{center}
\caption{Schematic bulk phase diagrams of binary Stockmayer fluid mixtures in the 
thermodynamic space of temperature $T$, pressure $p$, and chemical potential difference 
$\Delta\mu=\mu_B-\mu_A$ for various dipole moments and size parameters. The phase 
diagram in (a) represents a binary dipolar fluid mixture with the parameters  
$m_A^{\star}=0.75$, $m_B^{\star}=1.5$, and $\sigma_{BB}/\sigma_{AA}=1$. The topology of
this phase diagram is the same as that of a non-polar -- dipolar binary fluid mixture with 
the parameters $m_A^{\star}=0$, $m_B^{\star}=1.5$, 
and $\sigma_{BB}/\sigma_{AA}=1$ (compare Fig. \ref{fig8}(c)). Upon increasing the size of the 
component $B$ such that $\sigma_{BB}/\sigma_{AA}=2^{1/3}$ ($m_A^{\star}=0.75$, $m_B^{\star}=1.5$) 
one obtains the phase diagram shown in (c) in which the sheet $S_3$ appears again, which separates 
the B-rich isotropic liquid from the B-rich ferromagnetic liquid. The topology of this phase 
diagram is the same as that of non-polar -- dipolar binary fluid mixture with the parameters 
$m_A^{\star}=0$, $m_B^{\star}=1$, and $\sigma_{BB}/\sigma_{AA}=1$ 
(compare Fig. \ref{fig8}(b)). 
The phase diagram in (b) represents a binary dipolar fluid mixture with the parameters 
$m_A^{\star}=1$, $m_B^{\star}=2$, and $\sigma_{BB}/\sigma_{AA}=1$. The topology of this phase
diagram is similar to the topology of the phase diagram of non-polar -- dipolar binary 
fluid mixture with the parameters  $m_A^{\star}=0$, $m_B^{\star}=2$, and 
$\sigma_{BB}/\sigma_{AA}=1$ 
(compare Fig. \ref{fig8}(c)), with the difference that at the bottom of the 
winglike structure attached to the surface $S_1{\cup}S_2{\cup}S_3$ an A-rich 
ferromagnetic liquid phase appears. 
In this phase diagram a new sheet of second-order phase transition $S_3^{\prime}$ appears, 
which separates the A-rich isotropic liquid from the A-rich ferromagnetic liquid. 
The intersection of the winglike surface and the sheet $S_3^{\prime}$ 
forms a line $M_1$ of critical end points. The intersection of the sheets 
$S_1{\cup}S_2{\cup}S_3$ and $S_3^{\prime}$ forms another line $M_2$ of critical end points. 
Upon increasing the size of the component $B$ so that 
$\sigma_{BB}/\sigma_{AA}=2^{1/3}$ ($m_A^{\star}=1$, $m_B^{\star}=2$) one 
obtains the phase diagram shown in (d) where a winglike surface of first-order
phase transitions appears on the left part of the surface $S_1{\cup}S_2{\cup}S_3$. 
This winglike surface separates the vapor phase from the B-rich isotropic liquid phase.}
\label{fig16}
\end{figure}

\begin{figure}[t] 
\vspace{0.5cm}
\begin{center}
\rotatebox{0}{\scalebox{0.4}{\includegraphics{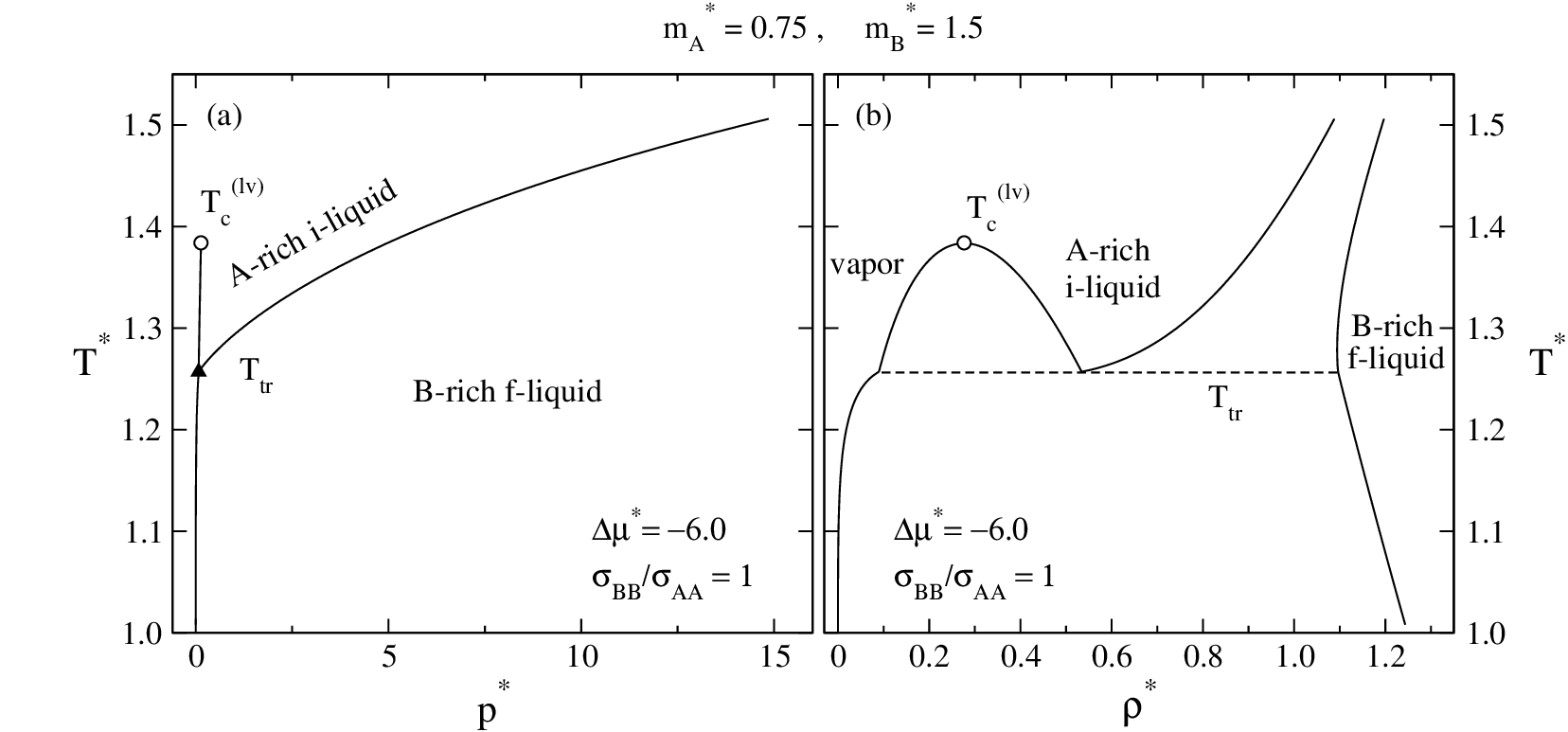}}}\\
\end{center}
\caption{Phase diagrams of a binary Stockmayer fluid mixture for dipole moments $m_A^{\star}=0.75$ and 
$m_B^{\star}=1.5$, and with size ratio $\sigma_{BB}/\sigma_{AA}=1$ in the pressure-temperature (a) and 
density-temperature (b) plane for $\Delta\mu^{\star}=-6$ (compare Fig. \ref{fig16}(a)).  
The LJ energy parameters are the same as in Fig. \ref{fig14}. 
Below the triple point temperature $T_{tr}$ the isotropic vapor coexists 
with the B-rich ferromagnetic liquid. Between $T_{tr}$ and $T_c^{(lv)}$ there are three phases: 
the vapor, the A-rich isotropic liquid, and the B-rich ferromagnetic liquid. The A-rich isotropic 
liquid -- B-rich ferromagnetic liquid coexistence does not end in a critical point within the range 
of fluid densities.}
\label{fig17}
\end{figure}

\begin{figure}[t] 
\vspace{0.5cm}
\begin{center}
\rotatebox{0}{\scalebox{0.4}{\includegraphics{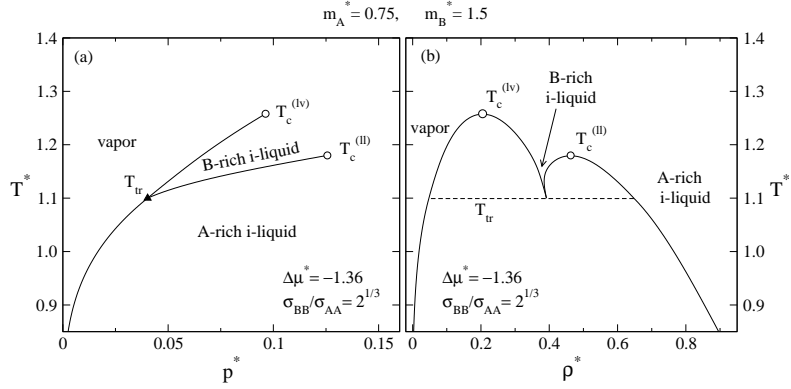}}}\\
\end{center}
\caption{Phase diagrams of a binary Stockmayer fluid mixture for dipole moments  $m_A^{\star}=0.75$ and 
$m_B^{\star}=1.5$, and with size ratio $\sigma_{BB}/\sigma_{AA}=2^{1/3}$  in the pressure-temperature 
(a) and density-temperature (b) plane for $\Delta\mu^{\star}=-1.36$ (compare Fig. \ref{fig16}(c)). 
The LJ energy parameters are the same as in Fig. \ref{fig14}. 
Below the triple point temperature the vapor coexists with the A-rich isotropic liquid. 
Between $T_{tr}$ and $T_c^{(ll)}$ there are three phases: the vapor, the 
B-rich isotropic liquid, and the A-rich isotropic liquid. 
Between $T_c^{(ll)}$ and $T_c^{(lv)}$ there is only vapor -- B-rich isotropic liquid coexistence.}
\label{fig18}
\end{figure}

\begin{figure}[t] 
\vspace{0.5cm}
\begin{center}
\rotatebox{0}{\scalebox{0.4}{\includegraphics{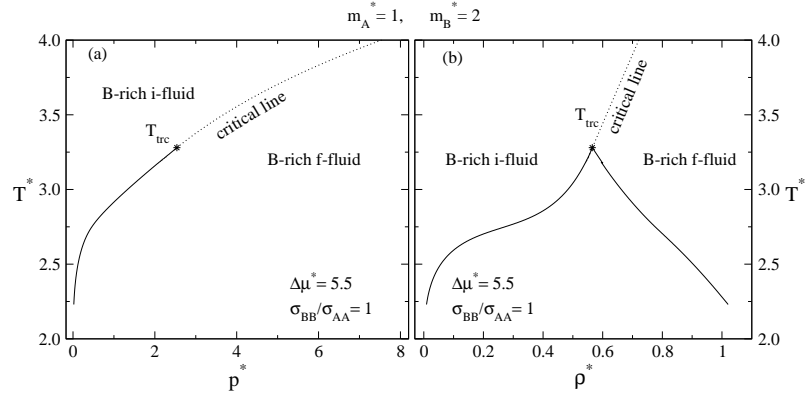}}}\\
\end{center}
\caption{Phase diagrams of a binary Stockmayer fluid mixture with dipole moments  
$m_A^{\star}=1$ and $m_B^{\star}=2$, and size ratio $\sigma_{BB}/\sigma_{AA}=1$ in 
the pressure-temperature (a) 
and density-temperature (b) plane for $\Delta\mu^{\star}=5.5$ (compare Fig. \ref{fig16}(b)). 
The LJ energy parameters are the same as in Fig. \ref{fig14}. 
Below the tricritical temperature $T_{trc}$ the B-rich isotropic fluid coexists with 
the B-rich ferromagnetic fluid. This first-order phase transition turns into a second-order 
phase transition at the tricritical temperature $T_{trc}$. Above  $T_{trc}$ the dotted 
line represents a line of second-order phase transitions.}
\label{fig19} 
\end{figure}

\begin{figure}[t] 
\vspace{0.5cm}
\begin{center}
\rotatebox{0}{\scalebox{0.4}{\includegraphics{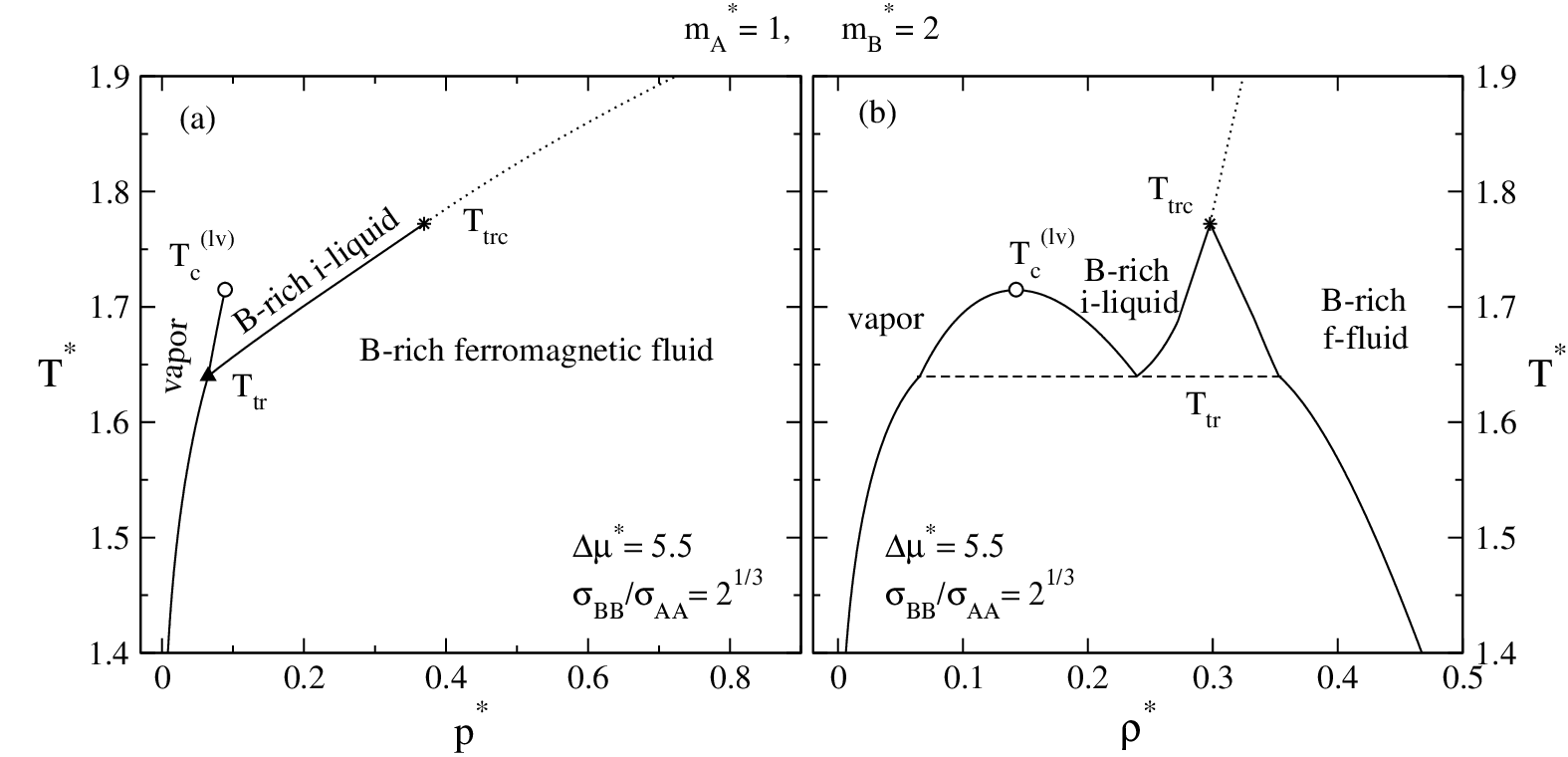}}}\\
\end{center}
\caption{Phase diagrams of binary Stockmayer fluid mixture for dipole moments  $m_A^{\star}=1$ and 
$m_B^{\star}=2$, and with size ratio $\sigma_{BB}/\sigma_{AA}=2^{1/3}$ in the pressure-temperature 
(a) and density-temperature (b) plane for $\Delta\mu^{\star}=5.5$ (compare Fig. \ref{fig16}(d)). 
The LJ energy parameters are the same as in Fig. \ref{fig14}. Below the triple point 
temperature $T_{tr}$ the isotropic vapor coexists with the B-rich ferromagnetic liquid. Between 
$T_{tr}$ and $T_c^{(lv)}$ there are three phases: the isotropic vapor, the B-rich isotropic 
liquid, and the B-rich ferromagnetic liquid. The line of first-order phase transition between 
the B-rich isotropic fluid and B-rich ferromagnetic liquid turns into a line of second-order 
phase transition (dotted line) at the tricritical temperature $T_{trc}$.}
\label{fig20} 
\end{figure}

Figure \ref{fig15} represents our numerical results for cuts through the phase diagrams of 
binary LJ mixtures for different size ratios $\sigma_{BB}/\sigma_{AA}$ and chemical potential 
differences $\Delta\mu^{\star}$. For the presentation we have changed the choice for 
$\Delta\mu^{\star}$ along with $\sigma_{BB}/\sigma_{AA}$ because there is no single value 
$\Delta\mu^{\star}$ for which the triple points exist for the different values 
$\sigma_{BB}/\sigma_{AA}$=1, 1.05, and $2^{1/3}$ (see Fig. \ref{fig14}). 
Figure \ref{fig15}(a) demonstrates that the liquid-vapor critical temperatures depend barely 
on the size ratio ($T_{c1}^{(lv)\star}=1.153$, $T_{c2}^{(lv)\star}=1.153$, and 
$T_{c3}^{(lv)\star}=1.152$) while the liquid-liquid critical temperatures exhibit a 
stronger size ratio dependence ($T_{c1}^{(ll)\star}=1.258$, $T_{c2}^{(ll)\star}=1.131$, 
and $T_{c3}^{(ll)\star}=1.148$). The corresponding critical densities are given by 
$\rho_{c1}^{(lv)\star}=0.273$, $\rho_{c2}^{(lv)\star}=0.217$, $\rho_{c3}^{(lv)\star}=0.195$ and
$\rho_{c1}^{(ll)\star}=0.708$, $\rho_{c2}^{(ll)\star}=0.502$, $\rho_{c3}^{(ll)\star}=0.453$. 
The corresponding critical pressures of the liquid-vapor and A-rich liquid -- B-rich liquid 
critical points are 
$p_{c1}^{(lv)\star}=0.113$, $p_{c2}^{(lv)\star}=0.091$, $p_{c3}^{(lv)\star}=0.083$ and      
$p_{c1}^{(ll)\star}=1.418$, $p_{c2}^{(ll)\star}=0.219$, $p_{c3}^{(ll)\star}=0.139$, 
respectively. 
Figure \ref{fig15} shows that the A-rich liquid -- B-rich liquid coexistence range 
within the pressure-temperature plane becomes narrower while within the density-temperature 
plane it grows wider with increasing size ratio.

We now turn to binary Stockmayer fluid mixtures whose components have different dipole moments 
and particle sizes. This model gives a reasonable description of bidisperse ferromagnetic fluids 
and as such provides also a first approximation of actual polydisperse ferrocolloids. 
For these fluids the magnetic dipole moments of the particles are proportional to the particle 
volume, i.e., $m_i$ $\propto$ $\sigma_{ii}^3$, $i=A,B$. In the following we restrict our study 
to binary Stockmayer fluid mixtures with dipole moment ratio $m_B^{\star}/m_A^{\star}=2$, which 
implies the size ratio $\sigma_{BB}/\sigma_{AA}=2^{1/3}$. This motivates our choice of the 
size ratio used for the above study of the phase behavior of the binary LJ fluid mixtures.

For the dipole moment ratio $m_B^{\star}/m_A^{\star}=2$ in the following we compare the 
phase diagrams of binary Stockmayer fluid mixtures composed of particles with either 
equal or different sizes. 
The corresponding schematic phase diagrams are shown in Fig. \ref{fig16}. 
Figure  \ref{fig16}(a) displays the schematic phase diagrams of a binary Stockmayer fluid mixture 
with dipole moments $m_A^{\star}=0.75$, $m_B^{\star}=1.5$, and with particle size ratio 
$\sigma_{BB}/\sigma_{AA}=1$; the LJ energy parameters are the 
same as in Fig. \ref{fig14}. 
The topology of this phase diagram is the same as that of shown in 
Fig. \ref{fig8}(c). This fact is not surprising because in both cases $m_B^{\star}=1.5$ and 
the increase of $m_A^{\star}$ from  $m_A^{\star}=0$ to $m_A^{\star}=0.75$ causes only a 
small change in the dipolar interaction energy relative to the LJ interaction energy.

Figure \ref{fig17} shows our numerical results for this binary Stockmayer fluid mixture with 
dipole moments  $m_A^{\star}=0.75$ and $m_B^{\star}=1.5$, and with particle size ratio 
$\sigma_{BB}/\sigma_{AA}=1$ for $\Delta\mu^{\star}=-6$. 
Figure \ref{fig17}(a) demonstrates that the plane $\Delta\mu^{\star}=-6$ cuts 
through the wing attached to the right hand side of the surface $S_1{\cup}S_2{\cup}S_3$ 
in Fig. \ref{fig16}(a). In accordance with the aforementioned topological similarity 
the phase diagrams shown in Fig. \ref{fig17} differ from those shown in 
Fig. \ref{fig13} only quantitatively. In Fig. \ref{fig17} the liquid-vapor 
critical point is given by  $T_c^{(lv)\star}=1.384$, $p_c^{(lv)\star}=0.138$, 
and $\rho_c^{(lv)\star}=0.277$. 
The corresponding triple point temperature and pressure are 
$T_{tr}^{\star}=1.257$ and $p_{tr}^{\star}=0.077$, respectively.

For the same dipole moments ($m_A^{\star}=0.75$ and $m_B^{\star}=1.5$) increasing the particle 
size ratio from $\sigma_{BB}/\sigma_{AA}=1$  to  $\sigma_{BB}/\sigma_{AA}=2^{1/3}$ 
the phase diagram shown in Fig. \ref{fig16}(a) changes considerably. 
The resulting phase diagram is shown in Fig. \ref{fig16}(c). The topology of this phase diagram 
is the same as in Fig. \ref{fig8}(b). 
As discussed in Subsec. III B the phase diagram shown in Fig. \ref{fig8}(b) emerges from 
the phase diagram shown in Fig. \ref{fig8}(c) by decreasing the dipole moment $m_B^{\star}$. 
Figure \ref{fig16}(a) and Fig. \ref{fig16}(c) show that the same topological change in the 
phase diagrams can be attained by increasing the corresponding particle size ratio. 
Figure \ref{fig16}(c) tells that the dipole moment strength $m_A^{\star}=0.75$ is insufficient 
to lead to the formation of an A-rich ferromagnetic liquid phase, so that for negative 
values of $\Delta\mu^{\star}$ the phase remains isotropic, as in the case of 
non-polar -- polar mixtures (see Fig. \ref{fig8}(b)). 

Figure \ref{fig18} shows our 
numerical findings for a binary Stockmayer fluid mixture with dipole moments 
$m_A^{\star}=0.75$ and $m_B^{\star}=1.5$, and with particle size ratio 
$\sigma_{BB}/\sigma_{AA}=2^{1/3}$ for $\Delta\mu^{\star}=-1.36$. Figure \ref{fig18}(a) 
demonstrates that the plane $\Delta\mu^{\star}=-1.36$ cuts through the upper part 
of the triple line $TL$ in Fig. \ref{fig16}(c) so that the corresponding triple 
point $T_{tr}$ describes a vapor -- A-rich isotropic liquid -- B-rich isotropic 
liquid coexistence with $T_{tr}^{\star}=1.1$ and $p_{tr}^{\star}=0.04$. 
The vapor-liquid and liquid-liquid critical points are given as 
$T_c^{(lv)\star}=1.258$, $p_c^{(lv)\star}=0.096$, $\rho_c^{(lv)\star}=0.204$ and 
$T_c^{(ll)\star}=1.18$, $p_c^{(ll)\star}=0.126$, $\rho_c^{(ll)\star}=0.462$, 
respectively. 
The topology of these phase diagrams is the same as those of shown in Fig. \ref{fig2}.

Increasing the dipole moments from $m_A^{\star}=0.75$ and $m_B^{\star}=1.5$ to 
$m_A^{\star}=1$ and $m_B^{\star}=2$ at a fixed particle size ratio 
$\sigma_{BB}/\sigma_{AA}=1$ the 
phase diagram shown in Fig. \ref{fig16}(a) changes considerably leading to the phase diagram 
displayed in Fig. \ref{fig16}(b). This increase of the dipole moments causes the 
disappearance of the winglike structure attached to the left part of the surface 
$S_1{\cup}S_2{\cup}S_3$ (see Fig. \ref{fig16}(b)), which is associated with the 
B-rich isotropic liquid -- vapor transitions. Moreover the winglike structure 
attached to the right part of the surface $S_1{\cup}S_2{\cup}S_3$ also changes due to 
the appearance of the sheet $S_3^{\prime}$. This sheet separates the A-rich isotropic 
liquid at high temperatures from the A-rich ferromagnetic liquid at low temperatures. 
$S_3^{\prime}$ is the locus of second-order phase transitions and is the remnant of the 
sheet $S_3$ (see Fig. \ref{fig1}(b)), which together with $S_1$ and $S_2$ has formed 
the surface $S_1{\cup}S_2{\cup}S_3$.

Figure \ref{fig19} displays our numerical results for a binary Stockmayer fluid mixture with 
dipole moments $m_A^{\star}=1$,  $m_B^{\star}=2$, and with particle size ratio 
$\sigma_{BB}/\sigma_{AA}=1$ for $\Delta\mu^{\star}=5.5$. 
For this value of $\Delta\mu^{\star}$ the figure shows a vertical slice through the left part 
of the surface $S_1{\cup}S_2{\cup}S_3$ in the phase diagram displayed in Fig. \ref{fig16}(b). 
The topology of these phase diagrams is the same as those of shown in Fig. \ref{fig11}. 
The tricritical point is given by $T_{trc}^{\star}=3.28$, $p_{trc}^{\star}=2.538$, and 
$\rho_{trc}^{\star}=0.565$.

Increasing the particle size ratio from $\sigma_{BB}/\sigma_{AA}=1$ to  
$\sigma_{BB}/\sigma_{AA}=2^{1/3}$ for fixed dipole moments $m_A^{\star}=1$ and 
$m_B^{\star}=2$, the phase diagram shown in Fig. \ref{fig16}(b) transforms into that 
shown in Fig. \ref{fig16}(d). Upon the increase of the particle size ratio on the left 
part of the surface $S_1{\cup}S_2{\cup}S_3$ in Fig. \ref{fig16}(d) a winglike structure 
appears, which is associated with B-rich isotropic liquid -- vapor phase transitions. 
Through a similar topological change, the phase diagram shown in Fig. \ref{fig8}(d) 
transforms into the phase diagram shown in Fig. \ref{fig8}(c) upon a decrease of the 
dipole moment $m_B^{\star}$  from $m_B^{\star}=1.5$ to $m_B^{\star}=1$.

Figure \ref{fig20} displays our numerical results for a binary Stockmayer fluid mixture with 
dipole moments $m_A^{\star}=1$, $m_B^{\star}=2$, and with a particle size ratio 
$\sigma_{BB}/\sigma_{AA}=2$ for $\Delta\mu^{\star}=5.5$. 
For this choice of $\Delta\mu^{\star}$ Fig. \ref{fig20}(a) represents a vertical slice 
through the left part of the surface $S_1{\cup}S_2{\cup}S_3$ in  the phase diagram shown 
in Fig. \ref{fig16}(d) where the attached winglike structure appears again. 
The corresponding liquid-vapor critical point, the 
tricritical point, and the triple point are given by  $T_c^{\star}=1.71$, $p_c^{\star}=0.09$, 
$\rho_c^{\star}=0.142$, $T_{trc}^{\star}=1.77$, $p_{trc}^{\star}=0.37$, 
$\rho_{trc}^{\star}=0.3$, and $T_{tr}^{\star}=1.64$, $p_{tr}^{\star}=0.065$, 
respectively. 
The topology of the phase diagrams shown in Fig. \ref{fig20} is the same as of 
those of shown in Fig. \ref{fig5}.       
 

\section{Summary}

In the present study of the phase behavior of binary Stockmayer 
(Eqs. (\ref{ljpot})-(\ref{totpot})) fluid mixtures 
the following main results have been obtained:

(1) We have extended the one-component version of the modified mean-field density functional
theory to that of binary Stockmayer fluid mixtures of particles A and B with  different dipole 
moments, sizes, and Lennard-Jones interaction energies. 
In particular, we have focused on the explicit formulation of the grand canonical 
functional, the equilibrium pressure $p$, and the chemical potentials $\mu_A$ and $\mu_B$ 
of the two components [see Subsec. II B and Eqs. (\ref{kem})-(\ref{press})].

(2) This system exhibits 6 distinct fluid phases: isotropic vapor, isotropic fluid, 
A-rich isotropic liquid, A-rich ferromagnetic liquid, B-rich isotropic liquid, and 
B-rich ferromagnetic liquid. At a given temperature $T$ the first-order phase transitions 
between them follow from the equalities of the chemical potentials and of the pressure between 
the coexisting phases [see Subsec. II C and Eq. (\ref{caca})].  
The loci of the second-order phase transitions between isotropic and ferromagnetic fluids 
can be determined analytically from a suitable Landau theory 
[see Subsec. II C and Eq. (\ref{sec2})].

(3) In the thermodynamic parameter space ($T,p,\Delta\mu=\mu_B-\mu_A$) we have constructed 
the schematic phase diagrams of binary fluid mixtures of particles with equal size and equal 
dipole moment (see Fig. \ref{fig1}). We have discussed the topological changes of these 
phase diagrams due to an increase of the particle dipole moments. The various shapes of 
the three-dimensional schematic phase diagrams are supported by numerically calculated 
cuts through them, which are two-dimensional phase diagrams at fixed chemical potential 
differences $\Delta\mu$. For dipole moments $m_A=m_B=0$ Fig. \ref{fig2} 
shows a cut through the phase coexistence surfaces (Fig. \ref{fig1}(a)) of first-order 
liquid-vapor and A-rich liquid -- B-rich liquid phase transitions of Lennard-Jones fluid 
mixtures which are characterized by energy parameters $\epsilon_{AA}$, $\epsilon_{BB}$, 
$\epsilon_{AB}$ and diameters $\sigma_{AA}$, $\sigma_{BB}$, 
$\sigma_{AB}=(\sigma_{AA}+\sigma_{BB})/2$. For reduced dipole moments 
$m_A^{\star}=m_B^{\star}=1$ with $m_i^{\star}=m_i/\sqrt{\epsilon_{AA}\sigma_{AA}^3}$, 
$i=A,B$, Fig. \ref{fig3} displays a cut through the phase coexistence  surfaces 
(Fig. \ref{fig1}(b)) describing the vapor-liquid, the vapor -- A-rich ferromagnetic liquid, 
and the A-rich isotropic liquid -- B-rich isotropic liquid phase transitions of binary 
Stockmayer fluid mixtures. For increased dipole moments $m_A^{\star}=m_B^{\star}=1.5$ 
the corresponding  schematic phase diagram is shown in Fig. \ref{fig1}(c). 
In this case Fig. \ref{fig4} shows the projections of the triple 
lines and critical lines onto a plane $p=const$. The vapor-liquid, vapor -- ferromagnetic 
fluid, isotropic fluid -- ferromagnetic fluid, 
and the A-rich ferromagnetic liquid -- B-rich ferromagnetic liquid  phase boundaries for 
dipole moments $m_A^{\star}=m_B^{\star}=1.5$ are shown in Figs. \ref{fig5} and \ref{fig6}. 
A further increase of the dipole moments to $m_A^{\star}=m_B^{\star}=2$ leads to the schematic 
phase diagram shown in  Fig. \ref{fig1}(d). A cut at $\Delta\mu^{\star}=const$ through 
those coexistence surfaces is shown in Fig. \ref{fig7}, describing the isotropic fluid 
-- ferromagnetic fluid and the A-rich ferromagnetic liquid -- B-rich 
ferromagnetic liquid phase transitions.

(4) We have also constructed the schematic phase diagrams of binary fluid mixtures of 
non-polar and various polar particles of equal size (see Fig. \ref{fig8}) and have 
discussed the topological changes of these phase diagrams due to an increase of the dipole 
moment of the polar component $B$. For dipole moments $m_A^{\star}=0$ and 
$m_B^{\star}=1$ Fig. \ref{fig9} shows a cut through the schematic phase diagram  
(Fig. \ref{fig8}(b)) describing the vapor-liquid, the A-rich isotropic 
liquid -- B-rich isotropic liquid, and the vapor -- ferromagnetic liquid phase coexistences. 
Figure \ref{fig10} displays another cut at $\Delta\mu^{\star}=const$ through 
the phase diagram in Fig. \ref{fig8}(b) indicating the possibility of A-rich isotropic 
liquid -- B-rich ferromagnetic liquid coexistence besides the vapor-liquid one. 
Upon increasing the dipole moment of the polar component $B$ from $m_B^{\star}=1$ to 
$m_B^{\star}=1.5$ the topology of the corresponding phase diagram shown in 
Fig. \ref{fig8}(b) changes to that shown in Fig. \ref{fig8}(c) which includes 
different types of phase coexistences: B-rich isotropic fluid -- B-rich ferromagnetic 
fluid coexistence (see Fig. \ref{fig11}), A-rich isotropic fluid -- B-rich ferromagnetic 
fluid coexistence (see Fig. \ref{fig12}), and vapor -- A-rich isotropic liquid 
coexistence together with A-rich isotropic liquid -- B-rich ferromagnetic liquid 
coexistence (see Fig. \ref{fig13}). A further increase of the dipole moment to  
$m_B^{\star}=2$ leads to the phase diagram shown in Fig. \ref{fig8}(d).

(5) We have studied the dependence of phase diagrams of binary Stockmayer 
liquid mixtures on the size ratio $\sigma_{BB}/\sigma_{AA}$. We have found 
that for  $1\leq\sigma_{BB}/\sigma_{AA}\leq2^{1/3}$ the phase diagrams of binary 
Lennard-Jones liquid mixtures ($m_A^{\star}=m_B^{\star}=0$) do not exhibit any topological 
change with increasing size ratio. For $m_A^{\star}=m_B^{\star}=0$ Figs. \ref{fig14} 
and \ref{fig15} demonstrate that with increasing particle size ratio only the 
thermodynamical parameter ranges of vapor-liquid and liquid-liquid coexistence are changed. 
In Fig. \ref{fig16} the schematic phase diagrams of binary Stockmayer liquid mixtures with 
particle size ratios $\sigma_{BB}/\sigma_{AA}=1$ and $\sigma_{BB}/\sigma_{AA}=2^{1/3}$ and 
for the dipole moment ratio $m_B^{\star}/m_A^{\star}=2$ are shown. 
Such binary mixtures with dipole moment and particle size ratios different from 1  can be 
considered as a first approximation of polydisperse ferrofluids. 
The topological change of the phase diagram of binary Stockmayer liquid mixtures with dipole 
moments $m_A^{\star}=0.75$ and  $m_B^{\star}=1.5$ due to the increase of the size ratio from 
$\sigma_{BB}/\sigma_{AA}=1$ to $\sigma_{BB}/\sigma_{AA}=2^{1/3}$ is displayed in  
Figs. \ref{fig16}(a) and  \ref{fig16}(c). Corresponding slices through these phase 
diagrams for $\Delta\mu^{\star}=const$ have been calculated (see Figs. \ref{fig17} 
and \ref{fig18}). Figure \ref{fig17} shows vapor -- A-rich isotropic liquid and A-rich 
isotropic liquid -- B-rich ferromagnetic liquid coexistence, while Fig. \ref{fig18} displays 
vapor -- B-rich isotropic liquid and B-rich isotropic liquid -- A-rich isotropic liquid 
coexistence. For dipole moments $m_A^{\star}=1$ and  $m_B^{\star}=2$ Figs. \ref{fig16}(b) 
and  \ref{fig16}(d) show another type of topological change in the phase diagrams of binary 
Stockmayer liquid mixtures due to the increase of the size ratio from 
$\sigma_{BB}/\sigma_{AA}=1$ to $\sigma_{BB}/\sigma_{AA}=2^{1/3}$. 
The corresponding slices through these phase diagrams for $\Delta\mu^{\star}=const$ 
have been calculated (see Figs. \ref{fig19} and \ref{fig20}). Figure \ref{fig20} shows that 
with increasing size ratio vapor-liquid coexistence emerges besides the coexistence 
of B-rich isotropic fluid and B-rich ferromagnetic fluid (Fig. \ref{fig19}).
     



\begin{thebibliography}{100}

\bibitem{teix1} P. I. C. Teixeira, J. M. Tavares, and M. M. Telo da Gama, 
J. Phys.: Condens. Matter {\bf 12}, R411 (2000).

\bibitem{groh1} B. Groh and S. Dietrich, in {\em New Approaches to Old and New Problems in 
Liquid State Theory - Inhomogeneities and Phase Separation in Simple, Complex and
Quantum Fluids}, proceedings of the NATO-ASI (Series C) held in Patti Marina (Messina),
Italy, July 7-17, 1998, edited by C. Caccamo, J. P. Hansen, and G. Stell (Kluwer, 
Dordrecht, 1999), Vol. C {\bf 529}, p. 173. 

\bibitem{tlusty} T. Tlusty and S. A. Safran, Science {\bf 290}, 1329 (2000);
S. A. Safran, Nature Materials {\bf 2}, 71 (2003).

\bibitem{egg} J. Eggebrecht, S. M. Thompson, and K. E. Gubbins, J. Chem. Phys. 
{\bf 86}, 2299 (1987).

\bibitem{teix2} P. I. C. Teixeira and M. M. Telo da Gama, J. Phys.: Condens. Matter 
{\bf 3}, 111 (1991).

\bibitem{frodl} P. Frodl and S. Dietrich, Phys. Rev. E {\bf 48}, 3741 (1993).

\bibitem{dang} L. X. Dang and T.-M. Chang, J. Chem. Phys. {\bf 119}, 9851 (2003).

\bibitem{enders} S. Enders, H. Kahl, M. Mecke, and J. Winkelmann, J. Mol. Liquids {\bf 115}, 
29 (2004). 

\bibitem{muk1} A. Mukhopadhyay, C. L. Caylov, and B. M. Law, Phys. Rev. E
{\bf 61}, R1036 (2000).

\bibitem{muk2} A. Mukhopadhyay and B. M. Law, Phys. Rev. E {\bf 63}, 011507 (2000).
\bibitem{cho1} J.-H. J. Cho and B. M. Law, Phys. Rev. Lett. {\bf 89}, 146101 (2002).
\bibitem{cho2} J.-H. J. Cho and B. M. Law, Phys. Rev. E {\bf 67}, 031605 (2003).
\bibitem{teix3} P. I. C. Teixeira, B. S. Almeida, M. M. Telo da Gama, J. A. Rueda, and
R. G. Rubio, J. Phys. Chem. {\bf 96}, 8488 (1992).

\bibitem{brad} A. Bradbury, S. Menear, R. W. Chantrell,
J. Magn. Magn. Mater. {\bf 54-57}, 745 (1986).

\bibitem{chan} K.-Y. Chan, K. E. Gubbins, D. Henderson, and L. Blum,
Mol.Phys. {\bf 66}, 299 (1989).
\bibitem{leeu1} S. W. de Leeuw, in {\em Condensed Matter Theories}, 
edited by L. Blum and F. B. Malik
(Plenum, New York, 1993), Vol. 8, p. 485.

\bibitem{jiang}  S. Jiang and K. S. Pitzer, J. Chem. Phys. {\bf 102}, 7632 (1995).
\bibitem{gao} G. T. Gao, J. B. Woller, X. C. Zeng, and W. Wang, J. Phys.:
Condens. Matter {\bf 9}, 3349 (1997).
\bibitem{blair}  M. J. Blair and G. N. Patey, Phys. Rev. E {\bf 57}, 5682 (1998).
\bibitem{cabral} B. J. C. Cabral, J. Chem. Phys. {\bf 112}, 4351 (2000).

\bibitem{kristof} T. Krist\'{o}f and I. Szalai, Phys. Rev. E {\bf 68}, 041109 (2003).

\bibitem{kristof2} T. Krist\'{o}f, J. Liszi, and I. Szalai, Phys. Rev. E {\bf 69}, 062106 (2004).

\bibitem{sadus} R. J. Sadus, Mol. Phys. {\bf 87}, 979 (1996).

\bibitem{leeu2} S. W. de Leeuw, B. Smit, and C. P. Williams,
J. Chem. Phys. {\bf 93}, 2704 (1990).

\bibitem{moo} G. C. A. M. Mooij, S. W. de Leeuw, B. Smit, and C. P. Williams,
J. Chem. Phys. {\bf 97}, 5113 (1992).
\bibitem{muller1} A. M\"uller, J. Winkelmann, T. Boubl\'ik, and J. Fischer,
Mol. Phys. {\bf 78}, 121 (1993).

\bibitem{muller2} A. M\"uller, J. Winkelmann, and J. Fischer, Fluid
Phase. Equilibria {\bf 99}, 35 (1994).
\bibitem{kriebel1} C. Kriebel, A. M\"uller, J. Winkelmann, and J. Fischer, 
Mol. Phys. {\bf 87}, 151 (1996).
\bibitem{wang} Z. Wang and C. Holm, Phys. Rev. E {\bf 68}, 041401 (2003).
\bibitem{morris} G. P. Morriss and D. J. Isbister, Mol. Phys. {\bf 59}, 911 (1986).
\bibitem{lee} P. H. Lee and B. M. Ladanyi, J. Chem. Phys. {\bf 91}, 7063 (1989).
\bibitem{chen1} X. S. Chen, M. Kasch, and F. Forstmann, Phys. Rev. Lett. {\bf 67}, 2674 (1991).
\bibitem{chen2} X. S. Chen and F. Forstmann, Mol. Phys. {\bf 76}, 1203 (1992).
\bibitem{troh} K. N. Trohidou and J. A. Blackman, Phys. Rev. B {\bf 51}, 11521 (1995).
\bibitem{zubar} A. Y. Zubarev, J. Exp. Theor. Phys. {\bf 93}, 80 (2001).
\bibitem{kant1} S. S. Kantorovich, J. Magn. Magn. Mater. {\bf 258-259}, 471 (2003).
\bibitem{kant2} S. S. Kantorovich and A. O. Ivanov, J. Magn. Magn. Mater. {\bf
252}, 244 (2002).

\bibitem{ivanov2} A. O. Ivanov and S. S. Kantorovich, Phys. Rev. E {\bf 70}, 021401 (2004).

\bibitem{kriebel2} C. Kriebel, A. M\"uller, J. Winkelmann, and J. Fischer, Fluid Phase 
Equilibria {\bf 119}, 67 (1996).
\bibitem{woj} M. Wojcik and K. E. Gubbins, Mol. Phys. {\bf 51}, 951 (1984).
\bibitem{boda} D. Boda, B. Kalm\'ar, J. Liszi, and I. Szalai, J. Chem. Soc., Faraday 
Trans. {\bf 92}, 2709 (1996).
\bibitem{valisko} M. Valisk\'o, D. Boda, J. Liszi, and I. Szalai, Phys. Chem.
Chem. Phys. {\bf 3}, 2995 (2001).
\bibitem{ivanov} A. O. Ivanov, J. Magn. Magn. Mater. {\bf 154}, 66 (1996).
\bibitem{islam} M. F. Islam, K. H. Lin, D. Lacoste, T. C. Lubensky, 
and A. G. Yodh, Phys. Rev. E {\bf 67}, 021402 (2003).
\bibitem{fenz} W. Fenz and R. Folk, Phys. Rev. E {\bf 67}, 021507 (2003).
\bibitem{sena1} S. Senapati and A. Chandra, J. Chem. Phys. {\bf 112}, 10467 (2000).
\bibitem{sena2} S. Senapati and A. Chandra, Phys. Rev. E {\bf 62}, 1017 (2000).
\bibitem{spol} C. Sp\"oler and S. H. L. Klapp, J. Chem. Phys. {\bf 118}, 3628 (2003).
\bibitem{ferna} M. J. Fernaud, E. Lomba, C. Martin, D. Levesque, and J.-J. Weis, J. 
Chem. Phys. {\bf 119}, 364 (2003).
\bibitem{linse} P. Linse, J. Chem. Phys. {\bf 86}, 4177 (1987).
\bibitem{matsu1} M. Matsumoto, Y. Takaoka, and Y. Kataoka, J. Chem. Phys. {\bf
98}, 1464 (1993).
\bibitem{matsu2} M. Matsumoto, H. Mizukuchi, and Y. Kataoka,
J. Chem. Phys. {\bf 98}, 1473 (1993).

\bibitem{huke} B. Huke and M. L\"ucke, Rep. Prog. Phys. {\bf 67}, 1731 (2004). 

\bibitem{dietrich1} S. Dietrich and A. Latz, Phys. Rev. B {\bf 40}, 9204 (1989).
\bibitem{dietrich2} S. Dietrich and M. Schick, Surf. Sci. {\bf 382}, 178 (1997).

\bibitem{scheffler} F. Scheffler, P. Maass, J. Roth, and H. Stark, Eur. Phys. J. B 
{\bf 42}, 85 (2004).

\bibitem{varga} I. Varga, F. Kun, and K. F. P\'al, Phys. Rev. E {\bf 69}, 030501(R) (2004).

\bibitem{mangold} K. Mangold, J. Birk, P. Leiderer, and C. Bechinger, Phys. Chem. Chem. Phys. 
{\bf 6}, 1623 (2004).

\bibitem{range1} G. M. Range and S. H. L. Klapp, Phys. Rev. E {\bf 69}, 041201 (2004).
\bibitem{range2} G. M. Range and S. H. L. Klapp, Phys. Rev. E {\bf 70}, 031201 (2004). 
\bibitem{range3} G. M. Range and S. H. L. Klapp, Phys. Rev. E {\bf 70}, 061407 (2004).

\bibitem{barker} J. A. Barker and D. Henderson, J. Chem. Phys. {\bf 47}, 4714 (1967).
\bibitem{groh2} B. Groh and S. Dietrich, Phys. Rev. Lett. {\bf 72}, 2422 (1994).
\bibitem{groh3} B. Groh and S. Dietrich, Phys. Rev. E {\bf 50}, 3814 (1994).
\bibitem{mansoori} G. A. Mansoori, N. F. Carnahan, K. E. Starling, and T. W. Leland,
J. Chem. Phys. {\bf 54}, 1523, (1971).
\bibitem{grad} I. S. Gradshteyn and I. M. Ryzhik, in {\em Table of Integrals,
  Series, and Products}, edited by A. Jeffrey and D. Zwillinger (Academic
  Press, New York, 
2000), p. 463. 
\bibitem{scott} R. L. Scott and P. H. van Konynenburg, Discuss. Faraday
Soc. {\bf 49}, 87 (1970).
\bibitem{konynen} P. H. van Konynenburg and R. L. Scott, Philos. Trans. R. Soc. London, 
Ser. A {\bf 298}, 495 (1980).
\end{thebibliography}
\end{document}